\documentstyle[seceq]{ptptex}
\title{%        %You can use \\ for explicit line-break
Description of $^8$Be as Deformed Gas-like \\ 
Two-Alpha-Particle States
}
\author{%       %Use \sc for the family name
Yasuro FUNAKI$^1$, Hisashi HORIUCHI$^1$, Akihiro TOHSAKI$^2$, 
\\ Peter SCHUCK$^3$, and Gerd R\"OPKE$^4$
}

\inst{%         %Affiliation, neglected when [addenda] or [errata]
{$^1$ Department of Physics, Kyoto University,
       Kyoto 606-8502, Japan}\\
  {$^2$ Department of Fine Materials Engineering, Shinshu 
        University,\\  Ueda 386-8567, Japan}\\
  {$^3$ Institut de Physique Nuceaire, 91406 Orsay Cedex, France}\\
  {$^4$ FB Physik, Universit\"at Rostock, D-18051 Rostock, 
       Germany}
}

\recdate{%      %Editorial Office will fill in this.
}

\abst{%       %this abstract is neglected when [addenda] or [errata]
In order to study non-zero spin excitations of the recently proposed $\alpha$-
cluster condensation in the self-conjugate 4n nuclei, spatial deformation is introduced into the model wave function of the $\alpha$-cluster condensate.  
The rotational band states of $^8$Be are investigated as a 
first step of a test case for the study of the deformation of the 
$\alpha$-cluster condensate.  Calculations reproduce well 
the binding energy of the $0^+$ ground state and also the excitation 
energy of the $2^+$ state.  Our $0^+$ wave function is found to 
be almost exactly equal to the $0^+$ wave function obtained by 
the generator coordinate method using Brink's 
$2\alpha$ wave function. The study shows that both the  $0^+$ 
ground and $2^+$ excited states can be considered as having a 
gas-like (i.e. weakly bound) 2$\alpha$-cluster structure. As for the  
$0^+$ ground state, the change of the wave function, due to the 
introduction of the deformation, is found to be very small.
}

\def\vc#1{\mbox{\boldmath $#1$}}
\begin{document}

\maketitle

\section{Introduction}
Recently we presented a conjecture that near the $n \alpha$ threshold 
in self-conjugate 4n nuclei there exist excited states of dilute 
density which are  composed of a weakly interacting gas of alpha 
particles and which can be considered as an $n \alpha$ condensed 
state\ \cite{thsr}. This conjecture was examined in $^{12}$C 
and $^{16}$O by using a new $\alpha$-cluster wave function of the 
$\alpha$-particle condensate type.  The second $0^+$ state 
in $^{12}$C near the $3 \alpha$ threshold was interpreted as such a 
condensed state as well as the fifth $0^+$ T=0 state in $^{16}$O near 
the $4 \alpha$-particle threshold. 

The new $\alpha$-cluster wave function of the $\alpha$-particle 
condensate type which we used in our previous study is an eigen 
state of zero angular momentum and represents a condensation of 
$\alpha$-particles in a spherically symmetric state.  Therefore this 
new $\alpha$-cluster wave function can describe only states of 
zero angular momentum. On the other hand, in $^{12}$C there exist 
some excited states above the $3 \alpha$ threshold which have 
non-zero angular momenta and are considered to have well-developed 
$3 \alpha$ cluster structure\ \cite{as}. Actually many microscopic 
$3 \alpha$ cluster model calculations gave the result that not 
only the second $0^+$ state near the $3 \alpha$ threshold but 
also some excited states with non-zero spin above the $3 \alpha$ 
threshold have $3 \alpha$ structure of dilute 
density\ \cite{supple}.  The question then arises whether we can 
identify some excitation modes of the $\alpha$-cluster 
condensate with  those excited states.

As a possible excitation mode of the $\alpha$-cluster condensate 
we can of course consider partial breaking of the condensate 
like that due to the promotion of one $\alpha$-cluster from the 
condensed $s$-orbit to a higher-lying orbit with non-zero spin. 
As another possibility for the excitation mode we can consider 
spatial deformation of the $\alpha$-cluster condensate. 
If the deformation is not energetically stable, we can expect 
vibrational excitations around the spherical shape, while if 
the deformation is stable, we can expect the appearance of 
rotational excitations.  

The purpose of the present paper is to study the possibility of 
spatial deformation of the $\alpha$-cluster condensate.  
This study is made in introducing a wave function of the 
$\alpha$-cluster condensate with deformation which is obtained 
by a natural modification of our previous wave function 
of the spherical $\alpha$-cluster condensate.  As a first step 
for the study of deformation, we investigate the $^8$Be nucleus. 
The ground state of $^8$Be is known to have a well-developed 
$2 \alpha$ cluster structure with large spatial separation 
between the two $\alpha$'s in relative $s$-state.  
With only two $\alpha$-particles this state can hardly be 
considered as an $\alpha$-particle condensation, but rather as 
a weakly bound state of two bosons in relative $s$-state thus exhibiting a gas-like structure of dilute density. Our wave function of 
Ref.\ \cite{thsr} is perfectly adapted to also treat only two 
$\alpha$'s and it will serve us as a test case to treat more 
$\alpha$-particles in the future.  The nucleus $^8$Be is known 
to have rotational excited states with large $\alpha$-decay widths 
$\Gamma_\alpha$, namely a $2^+$ state at 2.90 MeV with 
$\Gamma_\alpha$ = 1.45 MeV and a $4^+$ state at 11.4 MeV with 
$\Gamma_\alpha \sim $ 7 MeV\ \cite{be}. We will see that our 
investigation shows that in $^8$Be the $\alpha$-cluster condensed 
state is deformable and the experimentally observed rotational 
band can be explained as being generated from the $\alpha$-cluster 
condensate with deformation.  We will also see that our $0^+$ 
wave function is almost exactly equal to the $0^+$ wave function 
obtained by the generator coordinate method using 
Brink's $2\alpha$ wave function.

\section{Deformed $\alpha$-particle condensate}

\subsection{Wave function}

The wave function of the $\alpha$-cluster condensate with deformation 
can be expressed as follows by a slight
modification of our previous wave function of the spherical 
$\alpha$-cluster condensation: 
\begin{equation}
  |\Phi_{n\alpha} \rangle = (C_\alpha^\dagger)^n |{\rm vac} \rangle, 
\end{equation}
where the $\alpha$-particle creation operator $C_\alpha^\dagger$ 
in a deformed center of mass orbit is given by 
\begin{eqnarray}
  C_\alpha^\dagger &=& \int d^3 R\  
     \exp \bigl( - \frac{R_x^2}{\beta_x^2} - \frac{R_y^2}{\beta_y^2} - 
     \frac{R_z^2}{\beta_z^2} \bigr)
     \int d^3 r_1  \cdots d^3 r_4 \nonumber \\ 
     && \times \varphi_{0s}({\vc r}_1 - {\vc R}) 
     a_{\sigma_1 \tau_1}^\dagger ({\vc r}_1) 
     \cdots \varphi_{0s}({\vc r}_4 - {\vc R}) 
     a_{\sigma_4 \tau_4}^\dagger ({\vc r}_4).
\end{eqnarray}
Here 
\begin{equation}
 \varphi_{0s}({\vc r}) = (\pi b^2)^{-3/4} \exp\bigl(-\frac{{\vc r}^2}
 {2 b^2}\bigr),
\end{equation}
and $a_{\sigma \tau}^\dagger({\vc r})$ is the creation operator 
of a nucleon with spin-isospin $\sigma \tau$ at the spatial 
point ${\vc r}$.  Our previous wave function of the spherical 
$\alpha$-cluster condensation is just the case of 
$\beta_x=\beta_y=\beta_z=R_0$ in Eq.(2.2).  
The total $n \alpha$ wave function therefore can be written as 
\begin{eqnarray}
 &&\langle {\vc r}_1 \sigma_1 \tau_1, \cdots {\vc r}_{4n} \sigma_{4n} 
 \tau_{4n} |\Phi_{n\alpha} \rangle \nonumber \\ 
 && \propto {\cal A}\ \big[ \exp \big\{-\sum_{i=1}^n \bigl(\frac{2 X_{ix}^2}{B_x^2} 
 + \frac{2 X_{iy}^2}{B_y^2} + \frac{2 X_{iz}^2}{B_z^2}\bigr) \big\} \ 
  \phi(\alpha_1) \cdots \phi(\alpha_n) \big],
\end{eqnarray}
where $B_k^2 = b^2 + 2\beta_k^2$, ($k=x,y,z$), and  
${\vc X}_i = (1/4) \sum_n^4 {\vc r}_{in}$ is the center-of-mass 
coordinate of the $i$th $\alpha$-cluster $\alpha_i$. The internal 
wave function of the $\alpha$-cluster $\alpha_i$ is 
$\phi(\alpha_i) \propto \exp [ - (1/2b^2) \sum_n^4 ({\vc r}_{in} - 
{\vc X}_i )^2 ]$. The operator ${\cal A}$ is the total 
antisymmetrizer. It is to be noted that the wave function of Eqs.
(2.1) and (2.4) expresses the state where $n$ $\alpha$-clusters occupy 
the same deformed center of mass orbit 
$\exp \{-(\frac{2}{B_x^2} X_{x}^2 + \frac{2}{B_y^2} X_{y}^2 + 
\frac{2}{B_x^2} X_{z}^2) \}$, 
while the intrinsic $\alpha$-particle wave function stays spherical. 
We can easily see that the total center-of-mass motion is separated 
out of the wave function of Eqs.(2.1) and (2.4) because of the relation,  
\begin{equation}
 \sum_{i=1}^n X_{ik}^2 = n X_{Gk}^2 + \sum_{i=1}^n 
 (X_{ik} - X_{Gk})^2, \ (k=x,y,z), \ \ 
 {\vc X}_G \equiv (1/n) \sum_{i=1}^n {\vc X}_i.   
\end{equation}

The wave function of the $\alpha$-cluster condensation with 
deformation expressed by Eqs.(2.1) and (2.4) is a superposition 
of Brink's $\alpha$-cluster model wave function, 
$\Phi^{\rm B}({\vc R}_1, \cdots, {\vc R}_n)$, 
\begin{eqnarray}
 &&\Phi_{n\alpha}(\beta_x,\beta_y,\beta_z) \equiv 
 \langle {\vc r}_1 \sigma_1 \tau_1, \cdots {\vc r}_{4n} \sigma_{4n} 
 \tau_{4n} |\Phi_{n\alpha} \rangle \nonumber \\ 
 && = \int\! d^3\! R_1\cdots d^3\! R_n\ 
     \exp\! \big\{ -\sum_{i=1}^n \bigl( \frac{R_{ix}^2}{\beta_x^2} + 
     \frac{R_{iy}^2}{\beta_y^2} + \frac{R_{iz}^2}{\beta_z^2}\bigr) \big\} 
     \ \Phi^{\rm B}({\vc R}_1, \cdots, {\vc R}_n), \\ 
 &&\Phi^{\rm B}({\vc R}_1, \cdots, {\vc R}_n) \equiv 
 \det \{ \varphi_{0s}({\vc r}_1 - {\vc R}_1) 
     \chi_{\sigma_1 \tau_1} \cdots \varphi_{0s}
     ({\vc r}_{4n} - {\vc R}_n) \chi_{\sigma_{4n} \tau_{4n}} \},
\end{eqnarray}
where $\chi_{\sigma \tau}$ is the nucleon spin-isospin wave function. 
The form of Eq.(2.6) is directly obtained from Eqs.(2.1) $\sim$ (2.3), 
and the form of Eq.(2.4) can be obtained from Eq.(2.6)  using the 
relation, 
\begin{eqnarray}
 &&\frac{1}{\sqrt{4!}} \det \{ \varphi_{0s}({\vc r}_1 - {\vc R}) 
 \chi_{\sigma_1 \tau_1} \cdots \varphi_{0s}({\vc r}_4 - {\vc R}) 
 \chi_{\sigma_4 \tau_4} \} \nonumber \\ 
 &&\quad = \bigl(\frac{4}{\pi b^2}\bigr)^{3/4} 
 \exp\big\{-\frac{2}{b^2} ({\vc X} - {\vc R})^2\big\} \phi(\alpha),
\end{eqnarray}
where ${\vc X} = (1/4)\sum_{i=1}^4 {\vc r}_i$. 
An $\alpha$-condensed wave function with good angular momentum is 
obtained by projecting out the angular momentum from the deformed 
$\alpha$-condensed wave function as 
\begin{eqnarray}
 &&\Phi_{n\alpha}^J(\beta_x,\beta_y,\beta_z) = \int d\Omega 
 D^{J*}_{M K}(\Omega) {\widehat R}(\Omega) \Phi_{n\alpha}(\beta_x,
 \beta_y,\beta_z) \nonumber \\ 
 && = \int d\Omega D^{J*}_{M K}(\Omega) \int d^3 R_1\ \cdots d^3 R_n\ 
 \exp \bigl( -\sum_{i=1}^n \sum_{k=x,y,z} \frac{R_{ik}^2}{\beta_k^2} \bigr) 
 \nonumber \\ 
 && \quad \quad \times \Phi^{\rm B}(R^{-1}(\Omega){\vc R}_1, \cdots, 
 R^{-1}(\Omega){\vc R}_n), \nonumber \\ 
 && = \int d\Omega D^{J*}_{M K}(\Omega) \int d^3 R_1\ \cdots d^3 R_n\ 
 \exp \big\{ -\sum_{i=1}^n \sum_{k=x,y,z} \frac{(R(\Omega){\vc R}_i)_k^2}
 {\beta_k^2} \big\} \nonumber \\ 
 && \quad \quad \times \Phi^{\rm B}({\vc R}_1, \cdots, {\vc R}_n), 
\end{eqnarray}
where $\Omega$ is the Euler angle, $D^{J}_{M K}(\Omega)$ is the 
Wigner D-function, ${\widehat R}(\Omega)$ is the rotation operator, 
and $R(\Omega)$ is the 3 $\times$ 3 rotation matrix corresponding to 
${\widehat R}(\Omega)$. Here use is made of the relation, 
$\varphi_{0s}(R(\Omega){\vc r} - {\vc R}) = \varphi_{0s}
({\vc r} - R^{-1}(\Omega){\vc R})$.   It is to be noted that 
the rotation with respect to the total angular momentum is equivalent 
to the rotation with respect to the total orbital angular momentum, 
since the intrinsic spins of $\alpha$-clusters are saturated. 

In this paper we only treat the case of axially symmetric deformation 
taking the z-axis as the symmetry axis; 
$\beta_x = \beta_y \ne \beta_z$. In this case, we can replace 
$\int d\Omega$ by $\int d\cos \theta$, $D^{J}_{M K}(\Omega)$ by 
$d^{J}_{M 0}(\theta)$, ${\widehat R}(\Omega)$ by 
${\widehat R}_y(\theta)$, and $R(\Omega)$ by $R_y(\theta)$. 
Here ${\widehat R}_y(\theta)$ and $R_y(\theta)$ are the rotation 
operator and matrix, respectively, representing the rotation by 
the angle $\theta$ around the y-axis. 

\subsection{Elimination of spurious center-of-mass motion}

As already mentioned, the total center-of-mass motion is separated 
out of the wave function $\Phi_{n\alpha}(\beta_x,\beta_y,\beta_z)$. 
However, the separation of the total center-of-mass motion is no more 
true for the angular-momentum-projected wave function \\  
$\Phi^J_{n\alpha}(\beta_x,\beta_y,\beta_z)$.  Nevertheless the 
spurious admixture of the the total center-of-mass motion can be 
removed rather easily, as we explain now. 

By using Eq.(2.5) we separate the center-of-mass wave function from 
$\Phi_{n\alpha}(\beta_x,\beta_y,\beta_z)$ as 
\begin{eqnarray}
 \Phi_{n\alpha}(\beta_x=\beta_y,\beta_z) &=& Q \exp \big\{ -\frac{2n}
 {B_x^2} (X_{Gx}^2 + X_{Gy}^2) - \frac{2n}{B_z^2} X_{Gz}^2 \big\} \  
 {\widehat \Phi}_{n\alpha}(\beta_x=\beta_y,\beta_z), \nonumber  \\
 & & \\
 {\widehat \Phi}_{n\alpha}(\beta_x=\beta_y,\beta_z) &=& 
 {\cal A} \big[ \exp \big\{ - \sum_{i=1}^n \sum_{k=x,y,z} \frac{2}{B_k^2} 
 (X_{ik} - X_{Gk})^2 \big\} \  \phi(\alpha_1) \cdots \phi(\alpha_n) \big],\nonumber \\ 
  & &
\end{eqnarray}
where Q is some constant number.  Clearly 
${\widehat \Phi}_{n\alpha}(\beta_x=\beta_y,\beta_z)$ does not contain 
the center-of-mass coordinate ${\vc X}_G$ and is the internal wave 
function of $\Phi_{n\alpha}(\beta_x=\beta_y,\beta_z)$.  The correct 
$\alpha$-condensed wave function with good angular momentum is not 
$\Phi^J_{n\alpha}(\beta_x=\beta_y,\beta_z)$ given in Eq.(2.9) but is 
${\widehat \Phi}^J_{n\alpha}(\beta_x=\beta_y,\beta_z)$ defined as 
\begin{equation}
 {\widehat \Phi}^J_{n\alpha}(\beta_x=\beta_y,\beta_z) = \int d\cos 
 \theta \,d^J_{M0}(\theta) {\widehat R}_y(\theta) \ 
 {\widehat \Phi}_{n\alpha}(\beta_x=\beta_y,\beta_z). 
\end{equation}

The matrix element of a general translationally invariant scalar 
operator ${\widehat O}$ with the correct $\alpha$-condensed wave 
function ${\widehat \Phi}^J_{n\alpha}(\beta_x=\beta_y,\beta_z)$ 
is calculated as 
\begin{eqnarray}
 &&\frac {\langle {\widehat \Phi}^J_{n\alpha}(\beta_x=\beta_y,\beta_z) 
 |{\widehat O}|{\widehat \Phi}^J_{n\alpha}(\beta_x=\beta_y,\beta_z) 
 \rangle}{\langle {\widehat \Phi}^J_{n\alpha}(\beta_x=\beta_y,\beta_z) 
 |{\widehat \Phi}^J_{n\alpha}(\beta_x=\beta_y,\beta_z) \rangle} 
 \nonumber \\ 
 && \quad \quad = \frac{\int d\cos \theta \,d^J_{00}(\theta) \langle 
 {\widehat \Phi}_{n\alpha}(\beta_x=\beta_y,\beta_z)|{\widehat O}
 {\widehat R}_y(\theta)|{\widehat \Phi}_{n\alpha}(\beta_x=\beta_y,
 \beta_z) \rangle}{\int d\cos \theta \,d^J_{00}(\theta) \langle 
 {\widehat \Phi}_{n\alpha}(\beta_x=\beta_y,\beta_z)|
 {\widehat R}_y(\theta)|{\widehat \Phi}_{n\alpha}(\beta_x=\beta_y,
 \beta_z) \rangle}. 
\end{eqnarray}
The important technique for the calculation of the matrix elements 
of the integrands in this expression is to use the following relation 
\begin{eqnarray}
 &&\langle {\widehat \Phi}_{n\alpha}(\beta_x=\beta_y,\beta_z)|
 {\widehat O}{\widehat R}_y(\theta)|
 {\widehat \Phi}_{n\alpha}(\beta_x=\beta_y,\beta_z) \rangle 
 \nonumber \\ 
 &&\quad \quad = \frac{1}{Q^2 P(\theta)} \langle \Phi_{n\alpha} 
 (\beta_x=\beta_y,\beta_z)|{\widehat O}{\widehat R}_y(\theta)|
 \Phi_{n\alpha}(\beta_x=\beta_y,\beta_z) \rangle, \\
 && P(\theta) = \langle \exp \bigl( - \sum_{k=x,y,z} 
 \frac{2n}{B_k^2} X_{Gk}^2 \bigr) | {\widehat R}_y(\theta)| \exp \bigl( 
 - \sum_{k=x,y,z} \frac{2n}{B_k^2} X_{Gk}^2 \bigr) \rangle 
 \nonumber \\
 &&\quad \quad = \sqrt{ \bigl(\frac{\pi}{4n}\bigr)^3 \frac{B_x^6 B_z^4}
 {B_x^2 B_z^2 + (1/4)(B_x^2 - B_z^2)^2 \sin^2 \theta} }. 
\end{eqnarray}
This relation holds because the operator ${\widehat O}$ does not 
depend on the center-of-mass coordinate ${\vc X}_G$. 
By using this relation we get 
\begin{eqnarray}
 &&\frac {\langle {\widehat \Phi}^J_{n\alpha}(\beta_x=\beta_y,\beta_z) 
 |{\widehat O}|{\widehat \Phi}^J_{n\alpha}(\beta_x=\beta_y,\beta_z) 
 \rangle}{\langle {\widehat \Phi}^J_{n\alpha}(\beta_x=\beta_y,\beta_z) 
 |{\widehat \Phi}^J_{n\alpha}(\beta_x=\beta_y,\beta_z) \rangle} 
 \nonumber \\ 
 && \quad \quad = \frac{\int d\cos \theta \,d^J_{00}(\theta) \langle 
 \Phi_{n\alpha}(\beta_x=\beta_y,\beta_z)|{\widehat O}
 {\widehat R}_y(\theta)|\Phi_{n\alpha}(\beta_x=\beta_y,\beta_z) 
 \rangle 
 / P(\theta)} {\int d\cos \theta \,d^J_{00}(\theta) \langle 
 \Phi_{n\alpha}(\beta_x=\beta_y,\beta_z)|{\widehat R}_y(\theta)
 |\Phi_{n\alpha}(\beta_x=\beta_y,\beta_z) \rangle/ P(\theta)}.\nonumber \\ 
 & & 
\end{eqnarray}

The explicit formulas of the matrix elements of the overlap, 
the kinetic energy, the two-body nuclear force, and the Coulomb 
force in the case of the $^8$Be system are given in the Appendix A. 
In Appendix B we will show that the integration over the Euler 
angle $\theta$ can be performed analytically.

\subsection{Non-zero spin states on the $\beta_x(=\beta_y)=\beta_z$ 
line}

On the $\beta_x(=\beta_y)=\beta_z$ line the intrinsic state 
${\widehat \Phi}_{n\alpha}(\beta_x=\beta_y,\beta_z)$ is spherical 
and it may seem that we do not get non-zero spin states by the angular 
momentum projection.  But we can construct the non-zero spin states 
even on the $\beta_x(=\beta_y)=\beta_z$ line using angular 
momentum projection by the following usual limiting procedure: 
We first construct the non-zero spin state 
${\widehat \Phi}_{n\alpha}^J(\beta_x=\beta_y,\beta_z)$ of Eq.(2.12) 
for $\beta_x (= \beta_y) \ne \beta_z$ by angular momentum 
projection.  Next we normalize this state obtaining the normalized 
state ${\widehat \Phi}_{n\alpha}^{{\rm N}J}(\beta_x=\beta_y,\beta_z)$, 
\begin{equation}
 {\widehat \Phi}_{n\alpha}^{{\rm N}J}(\beta_x=\beta_y,\beta_z) = 
 \frac{{\widehat \Phi}_{n\alpha}^J(\beta_x=\beta_y,\beta_z)}
 {||{\widehat \Phi}_{n\alpha}^J(\beta_x=\beta_y,\beta_z)||},  
\end{equation}
where $||\Phi||$ denotes the norm of $\Phi$, $||\Phi|| = 
\sqrt{\langle \Phi | \Phi \rangle}$. Since 
$||{\widehat \Phi}_{n\alpha}^{{\rm N}J}(\beta_x=\beta_y,\beta_z) || 
= 1$, there exists the limiting state 
\begin{equation}
 \lim_{\beta_x \rightarrow \beta_z} 
 {\widehat \Phi}_{n\alpha}^{{\rm N}J}(\beta_x=\beta_y,\beta_z). 
\end{equation}
This limiting state is the non-zero spin wave function we adopt 
on the $\beta_x(=\beta_y)=\beta_z$ line. 

Since $\lim_{\beta_x \rightarrow \beta_z} 
{\widehat \Phi}_{n\alpha}^J(\beta_x=\beta_y,\beta_z) = 0$, 
the use of ${\widehat \Phi}_{n\alpha}^{{\rm N}J}(\beta_x=\beta_y,
\beta_z)$ near the limit of $\beta_x = \beta_z$ is not easy with 
respect to numerical accuracy.  Therefore in our present study 
of the $^8$Be system, we calculated the matrix elements with 
${\widehat \Phi}_{2\alpha}^{{\rm N}J}(\beta_x=\beta_y,\beta_z)$ 
analytically. This was made possible by performing the angular 
momentum projection analytically. Then the matrix elements with 
${\widehat \Phi}_{2\alpha}^{{\rm N}J}(\beta_x=\beta_y,\beta_z)$ 
near and on the $\beta_x(=\beta_y)=\beta_z$ line were obtained 
without any difficulty in the numerical calculations.  
Details are explained in Appendix C. 

\subsection{Microscopic Hamiltonian}

The Hamiltonian $H$ we use in this paper is written as 
\begin{equation}  
 H = T - T_G + V_N + V_C, 
\end{equation}
where $T$ is the total kinetic energy operator, $T_G$ is the 
center-of-mass kinetic enegy operator, $V_N$ is the effective 
two-body nuclear force, and $V_C$ is the Coulomb force between 
protons.  Explicit forms of $T$ and $T_G$ are as follows 
\begin{equation}
 T = \sum_{i=1}^8 \frac{-\hbar^2}{2 m} \bigl( 
 \frac{\partial}{\partial {\vc r}_i} \bigr)^2, \quad \ 
 T_G = \frac{-\hbar^2}{16 m} \bigl( 
 \frac{\partial}{\partial {\vc X}_G} \bigr)^2.
\end{equation}
As the effective two-body nuclear force, we adopt the 
force No.1 of Volkov\ \cite{volkov} which has the following form,
\begin{equation}
V_N = \frac{1}{2}\sum_{i\neq j}^8
\{(1-M)-MP_\sigma P_\tau\}_{ij} \sum_{n=1}^2v_n 
\exp\bigl(\!\!-\frac{r_{ij}^2}{a_n^2}\bigr).  
\end{equation}
The Coulomb force is given as 
\begin{equation}
V_C = \frac{1}{2}\sum_{i\neq j}^8 \frac{1-\tau_{zi}}{2} 
\frac{1-\tau_{zj}}{2} \frac{e^2}{r_{ij}}. 
\end{equation}
In our previous study of the alpha 
condensation in $^{12}$C and $^{16}$O, we used an effective 
nuclear force named F1 which contains a three-nucleon force in 
addition to the two-nucleon force\ \cite{tohsaki}.  The reason for 
this was that in both nuclei we treated not only the 
$\alpha$-condensed state of dilute density but also the 
normal-density states including the ground state.  
Therefore we had to use a nuclear 
force which has reasonable density-dependent properties.  
On the other hand, in the present study we treat only the ground 
band states of $^8$Be, all of which are expected to have a similar 
value of low density.  Therefore we here use a simpler effective 
nuclear force than in the previous study.  

\section{Results}

\subsection{Energy surface of the intrinsic state}

We first calculate the binding energy for the spherical state 
of $^8$Be which has two parameters $b$ and 
$\beta_x=\beta_y=\beta_z=R_0$. In Fig.1 we give the 
contour map of the energy surface in the two parameter space, 
$b$ and $\beta_x=\beta_y=\beta_z$.  The binding energy $E_\alpha$ 
of the single alpha particle which is independent of the Majorana 
parameter of the force No.1 of Volkov takes its lowest value 
$E_\alpha$ = -27.08 MeV for the size parameter $b$ = 1.37 fm.  
As the strength of the Majorana parameter $M$ of the Volkov 
force in $^8$Be, we adopt $M$ = 0.56. It is because with this 
value of $M$, the lowest binding energy takes the value -54.33 MeV 
as seen in Fig.1 which is only about 0.17 MeV lower than the 
theoretical two-alpha threshold energy, $2 E_\alpha = - 54.16$ MeV, 
in agreement with the experimental binding energy from the 
two-alpha threshold energy, about + 0.1 MeV.  
The value of the $b$ parameter at the energy minimum  
is $b$ = 1.36 fm which is very close to but 
slightly smaller\footnote{This smaller $b$-value may be interpreted 
as an effect of the Pauli-principle:one way to minimise the 
repulsive effect of antisymmetrisation is that the clusters make 
themselves smaller. This is clearly seen in the case of a low 
density gas of deuterons as explained in Ref.\ \cite{lombardo}} 
than the $b$ value of the free alpha particle.
It is known that this combination of the values of $b$ and $M$ 
which are close to $b$= 1.37 fm and $M$=0.56, respectively, gives a 
very good reproduction of the $\alpha$-$\alpha$ scattering phase 
shifts within the framework of the resonating group 
method\ \cite{ttt}. 

The qualitative feature of this map is similar to the corresponding 
maps in $^{12}$C and $^{16}$O given in our previous 
paper\ \cite{thsr}. This map shows a valley running from the outer 
region with large value of $\beta_x=\beta_y=\beta_z > 10$ fm 
and $b \approx$ 1.36 fm. The valley has a saddle point at 
$\beta_x=\beta_y=\beta_z \approx 8.5$ fm.  
Beyond the saddle point the binding energy is 
almost equal to $2 E_\alpha$.  The height of the saddle point 
measured from the theoretical threshold energy, $2 E_\alpha$, 
is about 0.56 MeV. The reason of the appearance of the saddle point 
is the same in the cases of $^{12}$C and $^{16}$O  and is 
attributed to the increase of the Coulomb energy 
and kinetic energy towards the inward direction which around the saddle point region is not yet 
compensated by the gain in nuclear potential energy.  The minimum of the energy surface is located 
at $\beta_x=\beta_y=\beta_z  \approx$ 3 fm. 

The result that the $\alpha$-condensed wave function reproduces well 
the experimental binding energy implies that the ground state of 
$^8$Be can be considered as a 2$\alpha$-particle $s$-wave quasi-bound state with a "gas-like" structure of dilute density. By "gas-like" we mean that the state is weakly bound and spreads widely forming a dilute density state and constituent clusters move rather freely. 

By fixing the $b$ parameter to $b$ = 1.36 fm, which is the $b$ value 
at the energy minimum of Fig.1, we now calculate the energy 
of the deformed states.  Fig.2 is the contour map of 
the energy surface in the two parameter space, $\beta_x (=\beta_y)$ 
and $\beta_z$.  This figure shows that the energy minimum 
lies on the $\beta_x = \beta_z$ line, which means that the 
energy minimum is attained for the spherical shape.  However, it is well 
known that nuclear deformation can not be deduced from a 
calculation without angular momentum projection. 
That is, we often encounter the case where the energy of the 
angular-momentum-projected state has a minimum for non-zero 
deformation although the energy of the intrinsic state without 
angular momentum projection favors a spherical shape. 

\subsection{Energy surfaces of spin-projected states}

In Fig.3 we show the contour map of the energy surface corresponding 
to the spin-projected states $J^\pi=0^+$ in the two parameter space, 
$\beta_x (=\beta_y)$ and $\beta_z$.  Now we see that the 
energy minimum no more lies on the $\beta_x = \beta_z$ line but 
is located at $\beta_x (= \beta_y) \approx$ 1.8 fm and 
$\beta_z \approx$ 7.8 fm in the prolate region of 
the map. The value of the energy minimum of -54.45 MeV means 
that the energy gain from angular momentum projection is only 
about 0.12 MeV.  Reflecting this small energy gain, there is a 
valley with an almost flat bottom running through this 
energy minimum to the second energy minimum at 
$\beta_x (= \beta_y) \approx$ 4.1 fm and $\beta_z \approx$ 0.0 fm 
in the oblate region of the map.  The bottom line 
of this valley crosses the $\beta_x=\beta_z$ line 
at $\beta_x=\beta_z=$ 3 fm which is just the energy minimum 
of Fig.2 for the spherical $\alpha$-condensed state. 
The variation of the binding energy along the bottom line of the 
valley is less than 0.12 MeV. 

How much different are the spin-projected wave functions along 
the bottom line of the valley from one another ?  The squared 
overlap $|\langle {\widehat \Phi}_{\rm I}^{0^+}|
{\widehat \Phi}_{\rm II}^{0^+} \rangle|^2$ of the normalized 
wave function ${\widehat \Phi}_{\rm I}^{0^+}$ at the energy minimum and 
the normalized wave function ${\widehat \Phi}_{\rm II}^{0^+}$ at 
$\beta_x=\beta_z=$ 3 fm is about 0.98.  This implies that 
${\widehat \Phi}_{\rm I}^{0^+}$ is almost equivalent to 
${\widehat \Phi}_{\rm II}^{0^+}$. We also calculated the squared 
overlap $|\langle {\widehat \Phi}_{\rm I}^{0^+}|
{\widehat \Phi}_{\rm III}^{0^+} \rangle|^2$ of 
${\widehat \Phi}_{\rm I}^{0^+}$ at the 
energy minimum and the normalized wave function 
${\widehat \Phi}_{\rm III}^{0^+}$ at the second energy minimum. The value of $|\langle 
{\widehat \Phi}_{\rm I}^{0^+}|{\widehat \Phi}_{\rm III}^{0^+} 
\rangle|^2$ is about 0.99. This result implies that 
${\widehat \Phi}_{\rm I}^{0^+}$ in the prolate region is 
almost equivalent to ${\widehat \Phi}_{\rm III}^{0^+}$ in the 
oblate region. Hence we can say that the $0^+$ component 
contained in the deformed intrinsic state along the bottom 
line of the valley is almost unchanged and is almost equivalent to the spherical 
condensed state at $\beta_x=\beta_z=$ 3 fm which is the 
energy minimum of Fig.2. 

In Fig.4 (a) we show the squared overlap 
$O(\beta_x, \beta_z) = |\langle {\widehat \Phi}_{2\alpha}^{{\rm N}0^+} 
(\beta_x, \beta_z) | {\widehat \Phi}_{\rm I}^{0^+} \rangle |^2$ 
in the form of the contour map in the two parameter space, 
$\beta_x (=\beta_y)$ and $\beta_z$, 
where ${\widehat \Phi}_{2\alpha}^{{\rm N}0^+}(\beta_x, \beta_z)$ 
is the normalized wave function of 
${\widehat \Phi}_{2\alpha}^{0^+}(\beta_x=\beta_y, \beta_z)$ of 
Eq.(2.12). The feature of the contour map of Fig.4 (a) is rather 
similar to the contour map of Fig.3.

Now we show in Fig.5 the contour map of the energy surface for the 
spin-projected states to $J^\pi=2^+$ in the two parameter space, 
$\beta_x (=\beta_y)$ and $\beta_z$.  Unlike the $0^+$ case, we see 
no local minimum. The binding energy becomes deeper as $\beta_x$ 
or $\beta_z$ becomes larger.  Let us consider the region between 
two lines, $\beta_x + \beta_z \approx$ 3 fm and $\beta_x + \beta_z 
\approx$ 7 fm.  The line, $\beta_x + \beta_z \approx$ 3 fm, is 
just the beginning of the energetically steeply-rising region towards 
$\beta_x = \beta_z$ = 0. The line, $\beta_x + \beta_z \approx$ 7 
fm, is approximately the boarder of the interaction region of 
two alpha particles.  The region between the two lines, 
$\beta_x + \beta_z \approx$ 3 fm and $\beta_x + \beta_z \approx$ 7 
fm is rather flat in energy, that is the binding energy in this 
region lies approximately between -51.0 MeV and -51.8 MeV.  The height 
of this plateau region measured from the $0^+$  energy minimum, 
-54.45 MeV, is approximately between 2.65 MeV and 3.45 MeV. 
Although, without imposing the correct boundary condition of 
the resonance, we cannot make a definite statement, we can safely 
conjecture that we will have 
a $2^+$ resonance of the excitation energy between 2.65 MeV 
and 3.45 MeV.  This excitation energy is in good agreement 
with the experimental $2^+$ excitation energy at 2.9 MeV. 
The reason of our conjecture is as follows.  Let us make the 
strength of the Majorana exchange mixture $M$ slightly smaller, 
taking $M$= 0.56 instead of $M$= 0.54. The contour map of the 
$J^\pi = 2^+$ energy surface for $M$ = 0.54 is shown in Fig.6.  
The feature of the contour map of Fig.6 is very similar to that of 
Fig.5, but we now have a local minimum at $\beta_x (= \beta_y) 
\approx$ 3.9 fm and $\beta_z \approx$ 0.0 fm in the oblate 
region of the map.  The binding energy of this local minimum 
is -52.28 MeV.  We show in Fig.7 the contour map of the $J^\pi = 0^+$ 
energy surface for the same $M$ = 0.54. The feature of the contour 
map of Fig.7 is very similar to that of Fig.3 with $M$ = 0.56 and 
the energy minimum which is slightly deeper than that of Fig.3 is 
now at -55.19 MeV.  Then the excitation energy of the $2^+$ state 
is 2.91 MeV.  This value of the $2^+$ excitation energy is very 
close to the experimental $2^+$ excitation energy of 2.9 MeV. 

In Fig.4 (b) we show the squared overlap 
$O(\beta_x, \beta_z) = |\langle {\widehat \Phi}_{2\alpha}^{{\rm N}2^+} 
(\beta_x, \beta_z) | {\widehat \Phi}_{\rm I}^{2^+} \rangle |^2$ 
in the form of the contour map in the two parameter space, 
$\beta_x (=\beta_y)$ and $\beta_z$, 
where ${\widehat \Phi}_{2\alpha}^{{\rm N}2^+}(\beta_x, \beta_z)$ 
is the normalized wave function of 
${\widehat \Phi}_{2\alpha}^{2^+}(\beta_x=\beta_y, \beta_z)$ of 
Eq.(2.12), and ${\widehat \Phi}_{\rm I}^{2^+}$ is the normalized 
wave function of the energy minimum of Fig.6, 
$\beta_x (= \beta_y) \approx$ 3.9 fm and $\beta_z \approx$ 0.0 fm. 
This figure shows that the $2^+$ wave function is almost unchanged 
from ${\widehat \Phi}_{\rm I}^{2^+}$ along the valley 
running from this energy minimum in the oblate 
region deeply into the region of prolate deformation. Hence we can say 
that the $2^+$ component contained in the deformed intrinsic state 
along this bottom line of the valley is almost unchanged. 

We also checked the cases of $M < 0.54$ for the $2^+$-state and 
again found the existence of a local minimum. 
The excitation energy of the $2^+$ state 
at the local minimum was found to be around 2.9 MeV. 

From the above results for the $0^+$ and $2^+$ states, in using 
the deformed $2\alpha$-state, we reach two conclusions: 
(1) For the $0^+$ state, the introduction of the deformed 
$2\alpha$-state plays essentially no role and the 
description with the spherical $2\alpha$-state is justified.  
(2) The deformed $2\alpha$-state yields a nice description 
of the $2^+$ state, which gives us a new understanding about the 
character of the $2^+$ state. Namely the $2^+$ state can be considered 
to be a gas-like state of $2\alpha$-particles in a relative $d$-wave.  
The merit of the introduction of the deformed $2\alpha$-state is 
just this point.  That is, we have obtained the picture about 
$^8$Be that not only the ground state but also the excited $2^+$ 
state can be interpreted in terms of a gas-like $2\alpha$-structure.  
This conclusion encourages us to introduce the deformed wave 
function of the $\alpha$-condensation in $^{12}$C and heavier 
self-conjugate 4n nuclei for a future study of the non-zero spin 
excitations of $\alpha$-condensate states. 

Let us now study the $4^+$ state. Fig.8 shows the contour map of the 
energy surface corresponding to the spin-projected states of 
$J^\pi=4^+$ in the two parameter space, $\beta_x (=\beta_y)$ and 
$\beta_z$.  Like the $2^+$ case there is no local minimum. 
We checked the cases of smaller $M$ values than 0.56 including 
0.54 but we found no local minimun in all cases. In Fig.8 we 
see that the approximately flat region of the $4^+$ surface is 
slightly pulled into a smaller $\beta_x + \beta_z$ region compared 
to the case of the $2^+$ energy surface.  The binding energy in 
the region between the two lines, $\beta_x + \beta_z \approx$ 2 fm 
and $\beta_x + \beta_z \approx$ 4 fm, lies approximately between 
-42 MeV and -45 MeV, which in terms of the excitation energy is 
between 9.45 MeV and 12.45 MeV. Although this excitation energy 
region between 9.45 MeV and 12.45 MeV is near the experimantal 
$4^+$ excitation energy, 11.4 MeV, we need more careful calculation 
including the resonance boundary condition in order to get 
reliable conclusion about the $4^+$ resonance.  

In this context it is useful to mention the calculation of the 
energy curve of $^8$Be using Brink's 2-$\alpha$ wave function.  
The energy curve is the binding energy as a function of 
the inter-$\alpha$ distance $R$. The energy curve for the 
$2^+$ resonance has a local minimum 
inside the interaction region with $R < 6$ fm, while that for 
the $4^+$ resonance does not have any local minimum.  

\subsection{Comparison with the solution of the $2\alpha$ 
Hill-Wheeler equation}

The wave function of the $2\alpha$-condensed state belongs to the 
functional space spanned by the wave functions of the following 
form, 
\begin{equation}
 {\cal A} \{ \chi({\vc r}) \phi(\alpha_1) \phi(\alpha_2) \}, 
\end{equation}
with $\chi({\vc r})$ representing an arbitrary function.  
Here ${\vc r}$ is the relative coordinate between two $\alpha$ 
clusters, ${\vc r} = {\vc X}_1 - {\vc X}_2$. 
The diagonalization of the Hamiltonian within this functional space 
is accomplished just by solving the well-known equation of the 
resonating group method (RGM), 
\begin{equation}
 \langle \phi(\alpha_1) \phi(\alpha_2) | (H - E) | {\cal A} 
 \{ \chi({\vc r}) \phi(\alpha_1) \phi(\alpha_2) \} \rangle = 0. 
\end{equation}
The solution of this RGM equation is equivalent to the solution 
of the $2\alpha$ Hill-Wheeler equation, 
\begin{equation}
 \int d^3 R\ \langle \Phi^B\bigl(\frac{1}{2}{\vc S}, -\frac{1}{2}{\vc S}\bigr) 
 | (H - E) | \Phi^B\bigl(\frac{1}{2}{\vc R}, -\frac{1}{2}{\vc R}\bigr) \rangle 
 \  f({\vc R}) = 0.
\end{equation}
Our wave functions of 2$\alpha$-condensation are approximations 
to the RGM wave functions ${\cal A} \{ \chi({\vc r}) \phi(\alpha_1) 
\phi(\alpha_2) \}$ obtained by solving Eq.(3.2) in imposing 
wave functions of Gaussian shapes.  Since the numerical 
solution of the 2$\alpha$ RGM equation of Eq.(3.2) is not difficult 
with  present-day computers, one may ask 
why we discuss approximations to the RGM wave functions. 
The answer is of course obvious: Our study in this paper 
is to extract and elucidate the $\alpha$-condensation character 
of the $^8$Be states which the RGM study of $^8$Be until now 
has not clarified even though it has given numerical values of 
the $^8$Be wave functions.  Furthermore our present study of the 
$^8$Be states is made in the scope which is not restricted only 
to the $^8$Be states but extends to excited states of 
general self-conjugate 4n nuclei which are of dilute density and 
are expected to exist near the $n \alpha$ breakup threshold. 
For the self-conjugate 4n nuclei, heavier than $^8$Be, especially 
for their excited states with dilute density, it is no longer easy 
but very difficult, especially for $n \ge 4$, to solve the RGM 
equation of motion, 
\begin{equation}
 \langle \phi(\alpha_1) \cdots \phi(\alpha_n) | (H - E) | {\cal A} 
 \{ \chi({\vc \xi}_1, \cdots {\vc \xi}_{n-1}) \phi(\alpha_1) \cdots 
 \phi(\alpha_n) \} \rangle = 0, 
\end{equation}
where ${\vc \xi}_1, \cdots {\vc \xi}_{n-1}$ are the relative 
coordinates between $n \alpha$ clusters.

For $0^+$, we solved the $2\alpha$ Hill-Wheeler equation of Eq.(3.3) 
and compared the solution with our wave function 
${\widehat \Phi}_{\rm I}^{0^+}$.  The Hill-Wheeler equation was 
solved by discretizing the radial integration as follows
\begin{eqnarray}
 &&\sum_j \langle \Phi^{{\rm B}0^+}(R_i)|(H-E)|\Phi^{{\rm B}0^+}(R_j)
 \rangle \ f(R_j) = 0,  \nonumber \\ 
 &&\Phi^{{\rm B}0^+}(R) \equiv \int d^2{\widehat R} \ 
 \Phi^{\rm B}\bigl( \frac{1}{2}{\vc R}, -\frac{1}{2}{\vc R} \bigr), 
\end{eqnarray}
where ${\widehat R}$ is the polar angle of ${\vc R}$. 
The energy expectation value of $\Phi^{{\rm B}0^+}(R)$ 
as a function of $R$, $\langle \Phi^{{\rm B}0^+}(R)|H| 
\Phi^{{\rm B}0^+}(R) \rangle / \langle \Phi^{{\rm B}0^+}(R)| 
\Phi^{{\rm B}0^+}(R) \rangle$, takes its minimum 
at $R=$ 3.45 fm with the minimum energy -52.069 MeV. It should be 
noted that this value is higher than the energy (-54.448 MeV) of 
${\widehat \Phi}_{\rm I}^{0^+}$ by 2.379 MeV, which is a rather large 
value.  By adopting $R_j = 0.5 \times j$ fm with $j= 1 \sim 23$, we 
obtained -54.444 MeV as the lowest eigen energy, and the squared 
overlap of the corresponding wave function with our wave function 
${\widehat \Phi}_{\rm I}^{0^+}$ is 0.9973.  When we adopt one more 
mesh point as $R_j = 0.5 \times j$ fm with $j= 1 \sim 24$, we 
obtained -54.446 MeV as the lowest eigen energy, and the squared 
overlap of the corresponding wave function with our wave function 
${\widehat \Phi}_{\rm I}^{0^+}$ becomes 0.9980.  We can say that 
we have obtained almost converged value for the lowest eigen energy 
of the Hill-Wheeler equation.  In principle, the lowest eigen 
energy of the Hill-Wheeler equation of Eq.(3.5) should be lower than 
the energy of ${\widehat \Phi}_{\rm I}^{0^+}$.  But the above values 
of the lowest eigen energy are still slightly higher than 
the energy of ${\widehat \Phi}_{\rm I}^{0^+}$. From this result 
we can say that our wave function ${\widehat \Phi}_{\rm I}^{0^+}$ 
is almost exactly equal to the wave function of the lowest energy 
solution of the Hill-Wheeler equation. This conclusion is 
surprising but it clearly shows that our model wave function 
${\widehat \Phi}^{{\rm N}0^+}_{2\alpha}(\beta_x=\beta_y,\beta_z)$ 
is very much suited to the ground state of $^8$Be 
which is a threshold state with dilute density.  
We can say that this conclusion strongly supports 
our new picture of the $^8$Be structure as a very dilute gas-like 
structure of 2 $\alpha$-particles rather than a dumb-bell structure 
of 2 $\alpha$-particles.

\section{Discussion} 

Our study in this paper is based on the use of the spin-projected 
state ${\widehat \Phi}_{2\alpha}^J(\beta_x=\beta_y,\beta_z)$ 
obtained from the deformed intrinsic state 
${\widehat \Phi}_{2\alpha}(\beta_x=\beta_y,\beta_z)$. 
Our deformed intrinsic state 
${\widehat \Phi}_{2\alpha}(\beta_x=\beta_y,\beta_z)$ is characterized 
as having a gas-like $2\alpha$-structure,i.e. a weakly bound 
2$\alpha$-state.  However, in order 
to be able to impose the same character to the spin-projected 
state ${\widehat \Phi}_{2\alpha}^J(\beta_x=\beta_y,\beta_z)$, 
it is necessary that the spin-projected state 
${\widehat \Phi}_{2\alpha}^{{\rm N}J}(\beta_x=\beta_y,\beta_z)$ 
is contained in the intrinsic state 
${\widehat \Phi}_{2\alpha}^{\rm N}(\beta_x=\beta_y,\beta_z)$ 
with non-negligible amplitude, where 
${\widehat \Phi}_{2\alpha}^{{\rm N}J}(\beta_x=\beta_y,\beta_z)$ and 
${\widehat \Phi}_{2\alpha}^{\rm N}(\beta_x=\beta_y,\beta_z)$ are 
the normalized states of 
${\widehat \Phi}_{2\alpha}^J(\beta_x=\beta_y,\beta_z)$ and 
${\widehat \Phi}_{2\alpha}(\beta_x=\beta_y,\beta_z)$, respectively. 
Therefore we have calculated the magnitude of the squared 
overlap amplitude, 
$|\langle {\widehat \Phi}_{2\alpha}^{{\rm N}J}(\beta_x=\beta_y,\beta_z) 
| {\widehat \Phi}_{2\alpha}^{\rm N}(\beta_x=\beta_y,\beta_z) \rangle 
|^2$, for $J = 0 \sim 4$, and the results are given in 
Figs.9 (a) $\sim$ (c) in the form of the contour map in the 
two parameter space, $\beta_x(=\beta_y)$ and $\beta_z$.  
We see that the minimum-energy states 
${\widehat \Phi}_{\rm I}^{0^+}$ and ${\widehat \Phi}_{\rm I}^{2^+}$ 
have strong overlap with their respective instrinsic states.

The energy surfaces given in Figs.3 and 5 are fairly flat. 
But at the same time we see in Figs.4 (a) and (b) that the wave 
functions in the energetically flat region are very similar to 
one another.  Therefore the flatness of the energy surface does not 
imply the appearance of the low-energy excited states. Also the 
superposition of the wave functions by solving the Hill-Wheeler 
equation will result in only a slight change of the energy and 
wave function from the energy and wave function at the 
energy minimum of the energy surface, respectively. 

As explained in {\S} 2.3, the non-zero spin states on the 
$\beta_x = \beta_z$ line were constructed by the limiting procedure 
and the matrix elements of these limiting states were calculated 
analytically. As is seen in the maps of Figs. 5, 6, and 8, the 
contour lines cross the $\beta_x = \beta_z$ line smoothly with 
no singular behavior near the $\beta_x = \beta_z$ line.
In Fig.10 we show the binding energy of the $2^+$ state along the 
the $\beta_x = \beta_z$ line in the case of $M$ = 0.56. 

In Fig.4(a) we saw that the minimum-energy  $0^+$ wave function 
which is projected out from the prolate intrinsic state is almost 
the same as the second-minimum-energy $0^+$ wave function 
which is projected out from the oblate intrinsic state.  
Similarly we saw in Fig.4(b) that the $2^+$ wave function 
which gives the local energy minimum and which is projected out from the 
oblate intrinsic state is almost the same as the $2^+$ wave functions 
which are projected out from the prolate intrinsic states. 
These results look very strange at first sight. In order to check 
whether the prolate intrinsic state is  very different from the 
oblate intrinsic state or not, we calculated the overlap between 
the intrinsic wave function ${\widehat \Phi}_{\rm I}$ at the 
energy minimum for $0^+$ which has prolate deformation with the general 
intrinsic state ${\widehat \Phi}_{2\alpha}(\beta_x=\beta_y,\beta_z)$ 
including the intrinsic wave function of the second energy 
minimum which has oblate deformation. In Fig.11 we show the squared 
overlap $|\langle {\widehat \Phi}_{2\alpha}^{\rm N}(\beta_x,\beta_z) | 
{\widehat \Phi}_{\rm I}^{\rm N} \rangle|^2$ in the form of a contour 
map in the two parameter space, $\beta_x(=\beta_y)$ and $\beta_z$, 
where ${\widehat \Phi}_{2\alpha}^{\rm N}(\beta_x,\beta_z)$ is the 
normalized wave function of 
${\widehat \Phi}_{2\alpha}(\beta_x=\beta_y,\beta_z)$ and 
${\widehat \Phi}_{\rm I}^{\rm N}$ is the normalized wave function of 
${\widehat \Phi}_{\rm I}$.  We see that the overlaps 
between the prolate intrinsic state ${\widehat \Phi}_{\rm I}^{\rm N}$ 
and the oblate intrinsic states near the second  energy minimum 
are very small. Below, we  explain the reason why we have such 
seemingly strange results for the spin-projected states and indicate 
that this  is due to the simplicity or the 
smallness of the functional space of the 2$\alpha$ system. 
The intrinsic wave function of our model which has the form of Eq.(2.11) 
with $n=2$ is written as 
\begin{eqnarray}
 &&{\widehat \Phi}_{2\alpha}(\beta_x=\beta_y,\beta_z) = 
  {\cal A} \big\{ \exp \bigl(-\frac{r_x^2+r_y^2}{B_x^2} -\frac{r_z^2}{B_z^2}\bigr)
   \phi(\alpha_1) \phi(\alpha_2) \big\},  \nonumber \\
 && r_k \equiv X_{1k} - X_{2k}, \quad \quad (k=x,y,z).  
\end{eqnarray}
The relative wave function 
$\exp \{-(r_x^2+r_y^2)/B_x^2 - r_z^2/B_z^2 \}$ can be expanded as 
follows 
\begin{eqnarray}
 && \exp\bigl(-\frac{r_x^2+r_y^2}{B_x^2} -\frac{r_z^2}{B_z^2}\bigr) 
  =\exp\big\{-\bigl(\frac{2}{3B_x^2} + \frac{1}{3B_z^2}\bigr) r^2 -\bigl(\frac{1}{3B_z^2} 
  - \frac{1}{3B_x^2}\bigr)(2r_z^2 - r_x^2 - r_y^2) \big\} \nonumber \\
 && =\exp\big\{-\bigl(\frac{2}{3B_x^2} + \frac{1}{3B_z^2}\bigr) r^2 \big\} - 
  \bigl(\frac{1}{3B_z^2} - \frac{1}{3B_x^2}\bigr)\sqrt{\frac{16\pi}{5}} 
  r^2 \exp\big\{-\bigl(\frac{2}{3B_x^2} + \frac{1}{3B_z^2}\bigr) r^2 \big\} \  
  Y_{20}({\hat r}) \nonumber \\
 && + \cdots ,
\end{eqnarray}
where ${\hat r}$ is the polar angle of ${\vc r} = {\vc X}_1 - 
{\vc X}_2$.  The sign of the coefficient of the second term, 
$(1/3B_z^2 - 1/3B_x^2)$, is opposite for prolate and oblate 
deformations, and it makes the overlap between prolate and 
oblate intrinsic states small. However once we project out the 
good spin state, the situation changes strongly. 
For the $0^+$-state, the projected 
wave function has the same following form for the leading term 
\begin{equation}
  {\cal A} \big[ \exp\big\{-\bigl(\frac{2}{3B_x^2} + \frac{1}{3B_z^2}\bigr) r^2 \big\}
   \phi(\alpha_1) \phi(\alpha_2) \big],  
\end{equation}
irrespectively of  the deformation of its original intrinsic state. 
Similarly for $2^+$, the projected wave function has the same 
following form for the leading term 
\begin{equation}
  {\cal A} \big[ r^2 \exp\big\{-\bigl(\frac{2}{3B_x^2} + \frac{1}{3B_z^2}\bigr) r^2 \big\}
   \  Y_{20}({\hat r})\  \phi(\alpha_1) \phi(\alpha_2) \big],  
\end{equation}
irrespectively of the deformation of its original intrinsic state. 
Therefore the overlap between spin-projected wave functions 
cannot become small but rather becomes close to unity as long as the 
values of $(2/B_x^2 + 1/B_z^2)$ are close to one another, 
irrespectively of the deformations of their intrinsic states. 
We can say that the essence of our above analysis is as follows: 
In the 2$\alpha$ system the spin-projected wave function is 
governed by a Gaussian-packet-like function $\chi_J(r)$ of a 
single coordinate $r$,  the relative distance of the two $\alpha$'s.  
Thus as long as one global character such as 
$\langle \chi_J | r^2 | \chi_J \rangle $ is similar, the 
spin-projected wave function can not  differ so much, 
and there is no room for a  difference with the original 
deformation.  We see in summary that the origin of the seemingly 
strange results about the spin-projected wave functions comes from 
the simplicity or the smallness of the functional space of the 
2$\alpha$ system.  Hence this kind of things will not be possible 
to occur for 3$\alpha$ and heavier systems. 

\section{Conclusion}

In order to study non-zero-spin excitations of  $\alpha$-cluster 
condensation in the self conjugate 4n nuclei which we proposed in 
our previous paper, we introduced a spatial deformation into the 
$\alpha$-cluster condensate.  The wave function of the 
$\alpha$-cluster condensate with deformation is obtained by a 
slight and natural modification of our previous wave function of 
the spherical condensate of $\alpha$-clusters. As a first step 
and test case for the study of deformation, we, in this paper, 
investigated the $^8$Be nucleus in applying the wave function of 
the deformed $\alpha$-cluster condensate to the $2\alpha$-system. 
The $0^+$ ground state, the $2^+$ excited state at 2.9 MeV 
with $\Gamma_\alpha = 1.45$ Mev, and the $4^+$ excited state at 11.4 
MeV with $\Gamma_\alpha \approx 7$ MeV are known to constitute a 
rotational band whose intrinsic state has a $2 \alpha$ cluster 
structure with large inter-$\alpha$ distance. Our study in this 
paper gave us the following conclusions: 
(1) the spherical $\alpha$-condensed wave function reproduces 
well the binding energy of the ground state, which implies  
that the $0^+$ ground state can be considered as a $2\alpha$-particle 
quasi-bound state in relative $s$-wave with gas-like structure 
of dilute density.  (2) The extension of the model space so as to 
include the deformed structure gives us only a slight energy gain 
of about 0.12 MeV for the $0^+$ ground state. 
Corresponding to this small change of the binding energy, the change 
of the wave function due to the extension of the model space is very 
small, although the deformed intrinsic wave function before the spin 
projection to $J=0$ at the energy minimum is quite different 
from the spherical optimum wave function. Thus as far as the 
$0^+$ ground state is concerned, the introduction of the deformed 
state plays essentially no role. (3) Our $0^+$ wave function is 
found to be almost exactly equal to the $0^+$ wave function 
obtained by the generator coordinate method using 
Brink's $2\alpha$ wave function.  This result strongly supports 
our new picture of the $^8$Be structure that it is more a very dilute gas-like 
structure of $2\alpha$-particles rather than a dumb-bell 
structure of $2\alpha$-particles.  (4) The deformed wave 
function reproduces well the observed excitation energy at 2.9 MeV 
of the $2^+$ state.  This result gives us the picture  
that not only the ground state but also the excited $2^+$ 
state of $^8$Be can be considered as a $2\alpha$-state with a 
gas-like structure of dilute density. 
This conclusion encourages us to introduce the 
deformed wave function of the $\alpha$-condensation also for $^{12}$C 
and heavier self-conjugate 4n nuclei for a future study of  non-zero 
spin excitations of the $\alpha$-condensation.

It is worth mentioning that our $\alpha$-particle condensed 
wave function, introduced in Ref.[1], has so far reproduced 
quite accurately $0^+$-threshold states in $^{12}$C and $^{16}$O 
and the rotational states in $^8$Be, treated in this paper. 
The fact that these results are obtained without a single 
adjustable parameter is, we believe, quite remarkable and 
gives strong credit to the correctness of the physics contained 
in our wave function.

\section*{Acknowledgements}

One of the authors (Y. F.) would like to thank Dr. M. Kimura and 
Prof. Y. Fujiwara for helpful advices for the calculations.

\appendix
\section{Matrix elements} %Empty argument \section{} yields `Appendix'. 

The deformed condensed wave function of Eq.(2.6) is written for the 
2$\alpha$ system as 
\begin{equation}
 \Phi_{2\alpha}(\beta_x=\beta_y,\beta_z) 
  = \int d^3 R_1\ d^3 R_2\ 
     \exp \bigl( -\sum_{i=1}^2 \sum_{k=x,y,z} \frac{R_{ik}^2}{\beta_k^2} 
     \bigr) \ \Phi^{\rm B}({\vc R}_1, {\vc R}_2). \nonumber
\end{equation}
By transforming ${\vc R}_1$ and ${\vc R}_2$ into the center-of-mass 
vector ${\vc R}_G$ and the relative vector ${\vc R}$, 
\begin{equation}
 {\vc R}_1 = {\vc R}_G + \frac{1}{2} {\vc R}, \ \ {\vc R}_2 = 
 {\vc R}_G - \frac{1}{2} {\vc R}, \nonumber
\end{equation}
this wave function is rewritten as 
\begin{eqnarray}
 &&\Phi_{2\alpha}(\beta_x=\beta_y,\beta_z) 
  = 4!\ \bigl(\frac{4}{\pi b^2}\bigr)^{3/2} \int d^3 R_G\ 
  \exp \big\{ - \sum_{k=x,y,z} \frac{2}{\beta_k^2} R_{Gk}^2 - 
  \frac{4}{b^2} ( {\vc X}_G - {\vc R}_G )^2 \big\} \nonumber \\
 && \quad \times \int d^3 R\ 
     \exp \bigl( -\sum_{k=x,y,z} \frac{1}{2 \beta_k^2} R_{k}^2 \bigr) \ 
     {\cal A} \big[ \exp \big\{ -\frac{1}{b^2} ( {\vc r} - {\vc R} )^2 \big\} 
     \phi(\alpha_1) \phi(\alpha_2) \big],  \nonumber
\end{eqnarray}
where ${\vc X}_G = ({\vc X}_1 + {\vc X}_2)/2$ and ${\vc r} = {\vc X}_1 
- {\vc X}_2$, with ${\vc X}_i$ denoting the center-of-mass 
coordinate of the $i$-th $\alpha$ cluster, $\alpha_i$.   
Therefore the internal wave function, ${\widehat \Phi}_{2\alpha}
(\beta_x=\beta_y,\beta_z)$, of Eq.(2.11) is written as 
\begin{eqnarray}
 {\widehat \Phi}_{2\alpha}(\beta_x=\beta_y,\beta_z) &=& Q' \int d^3 R\ 
     \exp \bigl( -\sum_{k=x,y,z} \frac{1}{2 \beta_k^2} R_{k}^2 \bigr) 
     \nonumber \\ 
     && \times {\cal A} \big[ \exp \big\{ -\frac{1}{b^2} 
     ( {\vc r} - {\vc R} )^2 \big\} \phi(\alpha_1) \phi(\alpha_2) \big]. 
     \nonumber
\end{eqnarray}
Now we introduce the following wave function, 
$\Psi_{2\alpha}(\beta_x=\beta_y,\beta_z)$, 
\begin{eqnarray}
 \Psi_{2\alpha}(\beta_x=\beta_y,\beta_z) &=& 4!\ 
 \bigl(\frac{4}{\pi b^2}\bigr)^{3/2} \frac{1}{Q'} \exp \bigl( -\frac{4}{b^2} 
  X_G^2 \bigr) {\widehat \Phi}_{2\alpha}(\beta_x=\beta_y,\beta_z) 
  \nonumber \\
  &=& \int d^3 R\ \exp \bigl( -\sum_{k=x,y,z} \frac{1}{2 \beta_k^2} 
  R_{k}^2 \bigr)\  \Phi^{\rm B}\bigl( \frac{1}{2}{\vc R}, -\frac{1}{2}
  {\vc R} \bigr). \nonumber
\end{eqnarray}
It is easy to prove the following relation 
\begin{eqnarray}
 &&\frac {\langle {\widehat \Phi}^J_{2\alpha}(\beta_x=\beta_y,\beta_z) 
 |{\widehat O}|{\widehat \Phi}^J_{2\alpha}(\beta_x=\beta_y,\beta_z) 
 \rangle}{\langle {\widehat \Phi}^J_{2\alpha}(\beta_x=\beta_y,\beta_z) 
 |{\widehat \Phi}^J_{2\alpha}(\beta_x=\beta_y,\beta_z) \rangle} 
 \nonumber \\ 
 && \quad \quad = \frac{\int d\cos \theta \,d^J_{00}(\theta) \langle 
 \Psi_{2\alpha}(\beta_x=\beta_y,\beta_z)|{\widehat O}
 {\widehat R}_y(\theta)|\Psi_{2\alpha}(\beta_x=\beta_y,
 \beta_z) \rangle}{\int d\cos \theta \,d^J_{00}(\theta) \langle 
 \Psi_{2\alpha}(\beta_x=\beta_y,\beta_z)|
 {\widehat R}_y(\theta)|\Psi_{2\alpha}(\beta_x=\beta_y,
 \beta_z) \rangle}.  \nonumber 
\end{eqnarray}

Below we give the explicit formulas of the matrix elements of 
the overlap, the kinetic energy, the two-body nuclear force, 
and the Coulomb force in the 2$\alpha$ system.  We use below 
the notation $\Phi^{\rm B}({\vc R})$, 
\begin{equation}
 \Phi^{\rm B}({\vc R}) \equiv \Phi^{\rm B}\bigl( \frac{1}{2}{\vc R}, 
 -\frac{1}{2}{\vc R} \bigr).  \nonumber
\end{equation}

The formula of the overlap is as follows. 
\begin{eqnarray}
 \lefteqn{\langle\Psi_{2\alpha}(\beta_x=\beta_y,\beta_z)|
 \widehat{R}_y(\theta)|\Psi_{2\alpha}(\beta_x=\beta_y,\beta_z)
 \rangle} \nonumber \\ 
 & & \nonumber \\ 
 &=& \int d^3 S d^3 R\,\exp\big\{\!-\!\!\!\sum_{k=x,y,z}\!\!\!
 \frac{S_k^2+(R_y(\theta){\vc R})_k^2}{2\beta_k^2}\big\} \ 
 \langle\Phi^{\rm B}({\vc S})\,|\,\Phi^{\rm B}({\vc R})\rangle 
 \nonumber \\
 &=& (2\pi)^3\beta_x^4\beta_z^2\ (G_0 + G_1 + G_2), \nonumber \\
 && \nonumber \\
 &G_0& = \frac{2}{\sqrt{(2\alpha_1^2+1)\big\{(2\alpha_1^2+1)
 (2\alpha_2^2+1)+\gamma\big\}}}, \nonumber \\ 
 &G_1& = -\frac{8}{\sqrt{(\frac{1}{2}\alpha_1^2+1)(\frac{3}{2}
 \alpha_1^2+1)\big\{(\frac{1}{2}\alpha_1^2+1)(\frac{3}{2}
 \alpha_1^2+1)(\frac{1}{2}\alpha_2^2+1)(\frac{3}{2}\alpha_2^2+1)
 +\frac{\gamma}{4}\big\}}}, \nonumber \\ 
 &G_2& =  \frac{6}{(\alpha_1^2+1)^2(\alpha_2^2+1)}\ .   \nonumber
\end{eqnarray}
In the above relation, we introduced variables, $\alpha_1$, 
$\alpha_2$, and $\gamma$, defined as 
\begin{equation}
 \alpha_1 \equiv \beta_x/b=\beta_y/b, \ \ \  \alpha_2 \equiv 
 \beta_z/b,\ \ \  \gamma \equiv (\alpha_1^2-\alpha_2^2)^2\sin^2\theta. 
 \nonumber
\end{equation}

The formula of the kinetic energy is as follows. 
\begin{eqnarray}
 &&\langle\Psi_{2\alpha}(\beta_x=\beta_y,\beta_z)|(T-T_G)
 \widehat{R}_y(\theta)|\Psi_{2\alpha}(\beta_x=\beta_y,\beta_z)
 \rangle \nonumber \\ 
 & & \nonumber \\ 
 && =\int d^3 S d^3 R\,\exp\big\{\!-\!\!\!\sum_{k=x,y,z}\!\!\!
 \frac{S_k^2+(R_y(\theta){\vc R})_k^2}{2\beta_k^2}\big\}\ 
 \langle \Phi^{\rm B}({\vc S})|T-T_G|\Phi^{\rm B}({\vc R})\rangle 
 \nonumber \\ 
 & & \nonumber \\ 
 &&= (2\pi)^3\beta_x^4\beta_z^2 \frac{\hbar^2}{mb^2}
 \Biggl( \frac{21}{4}(G_0+G_1+G_2) \nonumber \\ 
 & & +\frac{G_0}{2}\,\Big[\frac{3\alpha_1^2+1}{2\alpha_1^2+1}
 -\frac{\alpha_1^2+\alpha_2^2+1}{(2\alpha_1^2+1)
 \{(2\alpha_1^2+1)(2\alpha_2^2+1)+\gamma\}}\Big] 
 \nonumber \\ 
 & & \nonumber \\ 
 & & -\frac{G_1}{8}\,\Big[\frac{(3\alpha_1^2\alpha_2^2-4)
 (\frac{3}{4}\alpha_1^2\alpha_2^2+\alpha_1^2+\alpha_2^2+1)}
 {(\frac{1}{2}\alpha_1^2+1)(\frac{3}{2}\alpha_1^2+1) 
 (\frac{1}{2}\alpha_2^2+1)(\frac{3}{2}\alpha_2^2+1)+
 \frac{\gamma}{4}}+\frac{4(\frac{3}{2}\alpha_1^4+3\alpha_1^2+1)}
 {(\frac{1}{2}\alpha_1^2+1)(\frac{3}{2}\alpha_1^2+1)}\Big] 
 \nonumber \\ 
 & & \nonumber \\ 
 & & -\frac{G_2}{2}\,\frac{2\alpha_1^2(\alpha_2^2\!+\!1)\!+
 \!\alpha_2^2(\alpha_1^2\!+\!1)}{(\alpha_1^2+1)(\alpha_2^2+1)} 
 \Biggr). \nonumber
\end{eqnarray}

The formula of the two-body nuclear force is as follows. 
\begin{eqnarray}
 &&\langle\Psi_{2\alpha}(\beta_x=\beta_y,\beta_z)|V_N 
 \widehat{R}_y(\theta)|\Psi_{2\alpha}(\beta_x=\beta_y,\beta_z)
 \rangle \nonumber \\ 
 & & \nonumber \\ 
 && = \int d^3 S d^3 R\,\exp\big\{\!-\!\!\!\sum_{k=x,y,z}\!\!\!
 \frac{S_k^2+(R_y(\theta){\vc R})_k^2}{2\beta_k^2}
 \big\}\ \langle\Phi^{\rm B}({\vc S})|V_N|\Phi^{\rm B}({\vc R})
 \rangle \nonumber \\
 & & \nonumber \\
 &&= (2\pi)^3\beta_x^4\beta_z^2 
 \sum_{n=1}^2v_n\bigl(\frac{a_n^2}{a_n^2+2b^2}\bigr)^{3/2}
 \nonumber \\ 
 & & \times \Biggl(4(X_d+X_e)\bigl(\frac{G_0}{2}+
 \frac{G_1}{4}+\frac{G_2}{6}\bigr) \nonumber \\ 
 & & +\frac{4X_d}
 {\sqrt{(2\alpha_1^2\!+\!1)(u_n\alpha_1^2\!+\!1)
 \{(2\alpha_1^2\!+\!1)(u_n\alpha_1^2\!+\!1)(2\alpha_2^2\!+\!1)
 (u_n\alpha_2^2\!+\!1)\!+\!(u_n\!-\!2)^2\frac{\gamma}{4}\}}} 
 \nonumber \\ 
 & & +\frac{4X_e}{\sqrt{(\frac{1}{2}
 \alpha_1^2\!+\!1)\{(u_n\!+\!\frac{3}{2})\alpha_1^2\!+\!1\}
 \big[(\frac{1}{2}\alpha_1^2\!+\!1)\{(u_n\!+\!\frac{3}{2})
 \alpha_1^2\!+\!1\}(\frac{1}{2}\alpha_2^2\!+\!1)\{(u_n\!+\!
 \frac{3}{2})\alpha_2^2\!+\!1\}\!+\!(u_n\!+\!1)^2\frac{\gamma}{4}
 \big]}} \nonumber \\ 
 & & +\frac{4(X_e-2X_d)}
 {\sqrt{(\frac{3}{2}\alpha_1^2\!+\!1)\{(u_n\!+\!\frac{1}{2})
 \alpha_1^2\!+\!1\}\big[(\frac{3}{2}\alpha_1^2\!+\!1)\{(u_n\!+
 \!\frac{1}{2})\alpha_1^2\!+\!1\}(\frac{3}{2}\alpha_2^2\!+\!1)
 \{(u_n\!+\!\frac{1}{2})\alpha_2^2\!+\!1\}\!+\!(u_n\!-\!1)^2
 \frac{\gamma}{4} \big]}} \nonumber \\ 
 & & -\frac{16(X_d+X_e)}
 {\sqrt{\big[(\alpha_1^2\!\!+\!\!1)\{(\frac{1}{2}u_n\!\!+\!\!1)
 \alpha_1^2\!\!+\!\!1\}\!\!-\!\!\frac{\alpha_1^4}{4}\big]
 \Bigl(\big[(\alpha_1^2\!\!+\!\!1)\{(\frac{1}{2}u_n\!\!+\!\!1)
 \alpha_1^2\!\!+\!\!1\}\!\!-\!\!\frac{\alpha_1^4}{4}\big]
 \big[(\alpha_2^2\!\!+\!\!1)\{(\frac{1}{2}u_n\!\!+\!\!1)
 \alpha_2^2\!\!+\!\!1\}\!\!-\!\!\frac{\alpha_2^4}{4}\big]
 \!\!+\!\!\frac{\gamma}{4}\Bigr)}} \nonumber \\ 
 & & +\frac{4(X_d-2X_e)}
 {\sqrt{(\alpha_1^2\!+\!1)\{(u_n\!+\!1)\alpha_1^2\!+\!1\}
 \big[(\alpha_1^2\!+\!1)\{(u_n\!+\!1)\alpha_1^2\!+\!1\}
 (\alpha_2^2\!+\!1)\{(u_n\!+\!1)\alpha_2^2\!+\!1\}\!+
 \!u_n^2\frac{\gamma}{4}\big]}} \nonumber \\ 
 & & +\frac{16(X_d+X_e)}
 {(\alpha_1^2\!+\!1)\{(\frac{1}{2}u_n\!+\!1)\alpha_1^2\!+\!1\}
 \sqrt{(\alpha_2^2\!+\!1)\{(\frac{1}{2}u_n\!+\!1)
 \alpha_2^2\!+\!1\}}} \Biggr). \nonumber \\ 
 & &  \nonumber
\end{eqnarray}
As explained in section {\S}2.4, since we make use of 
the Volkov interaction as the two-body force, we should note 
that $V_N$ reads as follows,
\begin{eqnarray}
V_N = \frac{1}{2}\sum_{i\neq j}^8
\{(1-M)-MP_\sigma P_\tau\}_{ij} \sum_{n=1}^2v_n 
\exp\bigl(\!\!-\frac{r_{ij}^2}{a_n^2}\bigr).  \nonumber
\end{eqnarray}
In the above formula we used the variables defined below,
\begin{eqnarray}
 X_d&=&8-10M,\ \ X_e=10M-2,\nonumber \\ 
 u_n&=&\frac{b^2}{a_n^2+2b^2}.  \nonumber
\end{eqnarray}

As for the Coulomb force, we express it as a superposition of 
Gaussian functions which takes the form,
\begin{eqnarray}
 V_C &=& \frac{1}{2}\sum_{i\neq j}^8 \frac{1-\tau_{zi}}{2} 
 \frac{1-\tau_{zj}}{2} \frac{e^2}{r_{ij}} \nonumber \\ 
 &=& \frac{1}{2}\sum_{i\neq j}^8 \frac{1-\tau_{zi}}{2} 
 \frac{1-\tau_{zj}}{2} \frac{2e^2}{\sqrt{\pi}}
 \int_0^\infty d\eta \ \exp(-r_{ij}^2 \eta^2).  \nonumber
\end{eqnarray}
Therfore, we can easily obtain the matrix elements by changing 
the variables used in $V_N$ in the following way,
\begin{eqnarray}
       &&  a_n^2 \longrightarrow \frac{1}{\eta^2}, \nonumber \\ 
 &&\sum_{n=1}^2v_n\bigl(\frac{a_n^2}{a_n^2+2b^2}\bigr)^{3/2} 
 \longrightarrow \frac{2e^2}{\sqrt{\pi}}
 \int_0^\infty \bigl(\frac{1}{1+2b^2\eta^2}\bigr)^{3/2}d\eta, 
 \nonumber \\ 
       &&  X_d \longrightarrow 2, \nonumber \\
       &&  X_e \longrightarrow -1. \nonumber  
\end{eqnarray}
It is convenient to change the integration variable from $\eta$ to 
a new variable $\xi$ defined as 
\begin{eqnarray}
\xi^2= \frac{2b^2\eta^2}{1+2b^2\eta^2},\ \ \ \sqrt{\frac{1}{2b^2}}
d\xi=\bigl(\frac{1}{1+2b^2\eta^2}\bigr)^{3/2}d\eta.
\end{eqnarray}
The range of integration over $\xi$ is from zero to one. 
The numerical integration over $\xi$ was performed by using 
the Gauss-Legendre method.

\section{Analytical calculation of spin projection}

Matrix elements with the angular-momentum-projected wave functions 
are obtained by integrating the formulas of Appendix A multiplied 
by $d^J_{00}(\theta)$ over the angle $\theta$. We should note that 
in the formulas of Appendix A only $\gamma=(\alpha_1^2-\alpha_2^2)^2
\sin ^2\theta$ depends on the angle $\theta$. When limited to the 
case of $0^+$, $2^+$, and $4^+$, $d^J_{00}(\theta)$ contains only 
$\cos ^2\theta$ and $\cos ^4\theta$ as the terms which depend on 
$\theta$. Therefore, the integrals for $0^+$, $2^+$, and $4^+$ 
are classified into the following three types, 
\begin{eqnarray}
 \int_{-1}^1 dx \frac{1}{\sqrt{A^2-x^2}} &=& 2\arcsin
 \bigl(\frac{1}{A}\bigr), \nonumber \\ 
 \int_{-1}^1 dx \frac{x^2}{\sqrt{A^2-x^2}} &=& -\sqrt{A^2-1}+A^2
 \arcsin\bigl(\frac{1}{A}\bigr), \nonumber \\ 
 \int_{-1}^1 dx \frac{x^4}{\sqrt{A^2-x^2}} &=& -\frac{1}{4}(2+3A^2)
 \sqrt{A^2-1}+\frac{3}{4}A^4\arcsin\bigl(\frac{1}{A}\bigr), 
\end{eqnarray}
in the case of the overlap and nuclear force terms, and into 
the following three types,
\begin{eqnarray}
 \int_{-1}^1 dx \frac{1}{(A^2-x^2)^{3/2}} &=& \frac{2}
 {A^2\sqrt{A^2-1}}, \nonumber \\
 \int_{-1}^1 dx \frac{x^2}{(A^2-x^2)^{3/2}} &=& \frac{2}
 {\sqrt{A^2-1}}-2\arcsin\bigl(\frac{1}{A}\bigr), \nonumber \\
 \int_{-1}^1 dx \frac{x^4}{(A^2-x^2)^{3/2}} &=& \sqrt{A^2-1}+
 \frac{2A^2}{\sqrt{A^2-1}}-3A^2\arcsin\bigl(\frac{1}{A}\bigr),
\end{eqnarray}
in the case of the kinetic terms. Here $x = \cos \theta$ and 
we note that $A$ always takes values more than unity.

\section{Limit of $\beta_x \rightarrow \beta_z$ 
for spin projected states}

We stated in {\S}2.3, that there exist non-zero spin states on 
the $\beta_x(= \beta_y)=\beta_z$ line.  They are defined by the 
limiting procedure of $\beta_x \rightarrow \beta_z$.  
We can obtain the matrix elements with these limiting states 
analytically by performing Taylor expansion with respect to 
$s \equiv \alpha_1^2-\alpha_2^2$ around $s=0$. In this appendix, 
we display the series expansions with respect to $s$ up to the 
second order for the case of the $2^+$ state. 
For the sake of the series expansion, we introduce 
$t \equiv \alpha_1^2 + \alpha_2^2$ and rewrite $\alpha_1^2$ and 
$\alpha_2^2$ by $s$ and $t$ as 
\begin{equation}
 \alpha_1^2 = \frac{1}{2} ( s + t ),\  \alpha_2^2 = 
 \frac{1}{2} ( t - s ). 
\end{equation}
By using 
\begin{equation}
 d^2_{00}(x) = \frac{3}{2}x^2 -\frac{1}{2},
\end{equation}
we get the series expansions about the overlap terms as follows,
\begin{eqnarray}
 \widehat{G}_0&=&\int_{-1}^1 dx \bigl(\frac{3}{2}x^2-\frac{1}{2}\bigr)\, G_0
 =\frac{4}{15}\frac{1}{(t+1)^{7/2}}s^2-\frac{2}{15}\frac{1}
 {(t+1)^{9/2}}s^3+\cdots,  \nonumber \\ 
 \widehat{G}_1&=&\int_{-1}^1 dx \bigl(\frac{3}{2}x^2-\frac{1}{2}\bigr)\, G_1
 =-\frac{4}{15}\frac{1}{\{(\frac{1}{4}t+1)
 (\frac{3}{4}t+1)\}^{7/2}}s^2 \nonumber  \\
 && \quad +\frac{1}{30}
 \frac{\frac{3}{2}t+4}{\{(\frac{1}{4}t+1)
 (\frac{3}{4}t+1)\}^{9/2}}s^3-\cdots, \nonumber  \\ 
 \widehat{G}_2&=&\int_{-1}^1 dx (\frac{3}{2}x^2-\frac{1}{2})\, G_2=0.
 \end{eqnarray}

About the kinetic terms, we get 
\begin{eqnarray}
 &&\int_{-1}^1dx \bigl(\frac{3}{2}x^2-\frac{1}{2}\bigr)
 \langle\Psi_{2\alpha}(\beta_x=\beta_y,\beta_z)|(T-T_G)
 \widehat{R}_y(x)|\Psi_{2\alpha}(\beta_x=\beta_y,\beta_z)
 \rangle \nonumber \\ 
 && \nonumber \\ 
 &&= (2\pi)^3\beta_x^4\beta_z^2 \frac{\hbar ^2}{mb^2} \Bigg[
 \frac{21}{4}(\widehat{G}_0+\widehat{G}_1+\widehat{G}_2) \nonumber \\ 
 && -\frac{1}{15}\frac{3t-4}{(t+1)^{9/2}}s^2
 +\frac{1}{10}\frac{t-2}{(t+1)^{11/2}}s^3+\cdots \nonumber \\ 
 && +\frac{1}{120}\frac{15t^2+24t-32}{\{(\frac{3}{4}t+1)
 (\frac{1}{4}t+1)\}^{9/2}}s^2
 -\frac{1}{640}\frac{15t^3+56t^2-128}{\{(\frac{3}{4}t+1)
 (\frac{1}{4}t+1)\}^{11/2}}s^3+\cdots  \nonumber \\
 && - 0\  \Bigg].
 \end{eqnarray}

About the terms which come from the nuclear force, we have  
\begin{eqnarray}
 &&\int_{-1}^1\bigl(\frac{3}{2}x^2-\frac{1}{2}\bigr)\langle
 \Psi_{2\alpha}(\beta_x=\beta_y,\beta_z)|V_N 
 \widehat{R}_y(x)|\Psi_{2\alpha}(\beta_x=\beta_y,\beta_z)
 \rangle \nonumber \\ 
 && \nonumber \\ 
 &&= (2\pi)^3\beta_x^4\beta_z^2 
 \sum_{n=1}^2v_n\bigl(\frac{a_n^2}{a_n^2+2b^2}\bigr)^{3/2} 
 \nonumber \\ 
 && \Bigg\{4(X_d+X_e)\bigl(\frac{\widehat{G}_0}{2}+
 \frac{\widehat{G}_1}{4}+
 \frac{\widehat{G}_2}{6}\bigr) \nonumber \\ 
 && +4X_d\Biggl(\frac{4}{15}\frac{\sqrt{2}(u_n-2)^2}
 {\big\{(t+1)(u_nt+2)\big\}^{7/2}}s^2
 -\frac{2}{15}\frac{\sqrt{2}(u_n-2)^2(2u_nt+u+2)}{\big\{(t+1)
 (u_nt+2)\big\}^{9/2}}s^3+\cdots \Biggr) \nonumber \\ 
 && +4X_e\Biggl( \frac{1}{30}\frac{(u_n+1)^2}
 {\big\{(\frac{1}{4}t+1)(\frac{2u_n+3}{4}t+1)\big\}^{7/2}}s^2
 -\frac{1}{120}\frac{(u_n+1)^2\bigl(\frac{2u_n+3}{4}t+u_n+2\bigr)}
 {\big\{(\frac{1}{4}t+1)(\frac{2u_n+3}{4}t+1)\big\}^{9/2}}s^3+
 \cdots \Biggr) \nonumber \\
 && +4(X_e-2X_d)\Biggl( \frac{1}{30}\frac{(u_n-1)^2}
 {\big\{(\frac{3}{4}t+1)(\frac{2u_n+1}{4}t+1)\big\}^{7/2}}s^2
 -\frac{1}{120}\frac{(u_n-1)^2\bigl(\frac{6u_n+3}{4}t
 +u_n+2\bigr)}{\big\{(\frac{3}{4}t+1)(\frac{2u_n+1}{4}t+1)
 \big\}^{9/2}}s^3+\cdots \Biggr) \nonumber \\
 && -16(X_d+X_e)\Biggl( \frac{1}{30}\frac{1}{\bigl
 (\frac{2u_n+3}{16}t^2+\frac{u_n+4}{4}t+1\bigr)^{7/2}}s^2
 -\frac{1}{120}\frac{\frac{2u_n+3}{4}t+\frac{u_n}{2}+2}{\bigl
 (\frac{2u_n+3}{16}t^2+\frac{u_n+4}{4}t+1\bigr)^{9/2}}s^3+\cdots 
 \Biggr) \nonumber \\ 
 && +4(X_d-2X_e)\Biggl( \frac{64}{15}\frac{u_n^2}{\big[(t+2)
 \{(u_n+1)t+2\}\big]^{7/2}}s^2
 -\frac{64}{15}\frac{\{(u_n+1)t+u_n+2\}u_n^2}{\big[(t+2)
 \{(u_n+1)t+2\}\big]^{9/2}}s^3+\cdots \Biggr) \nonumber \\ 
 && +16(X_d+Xe)\times 0 
 \Bigg\}. 
\end{eqnarray}

\end{document}

\begin{figure}
\epsfxsize = \hsize
\epsfbox{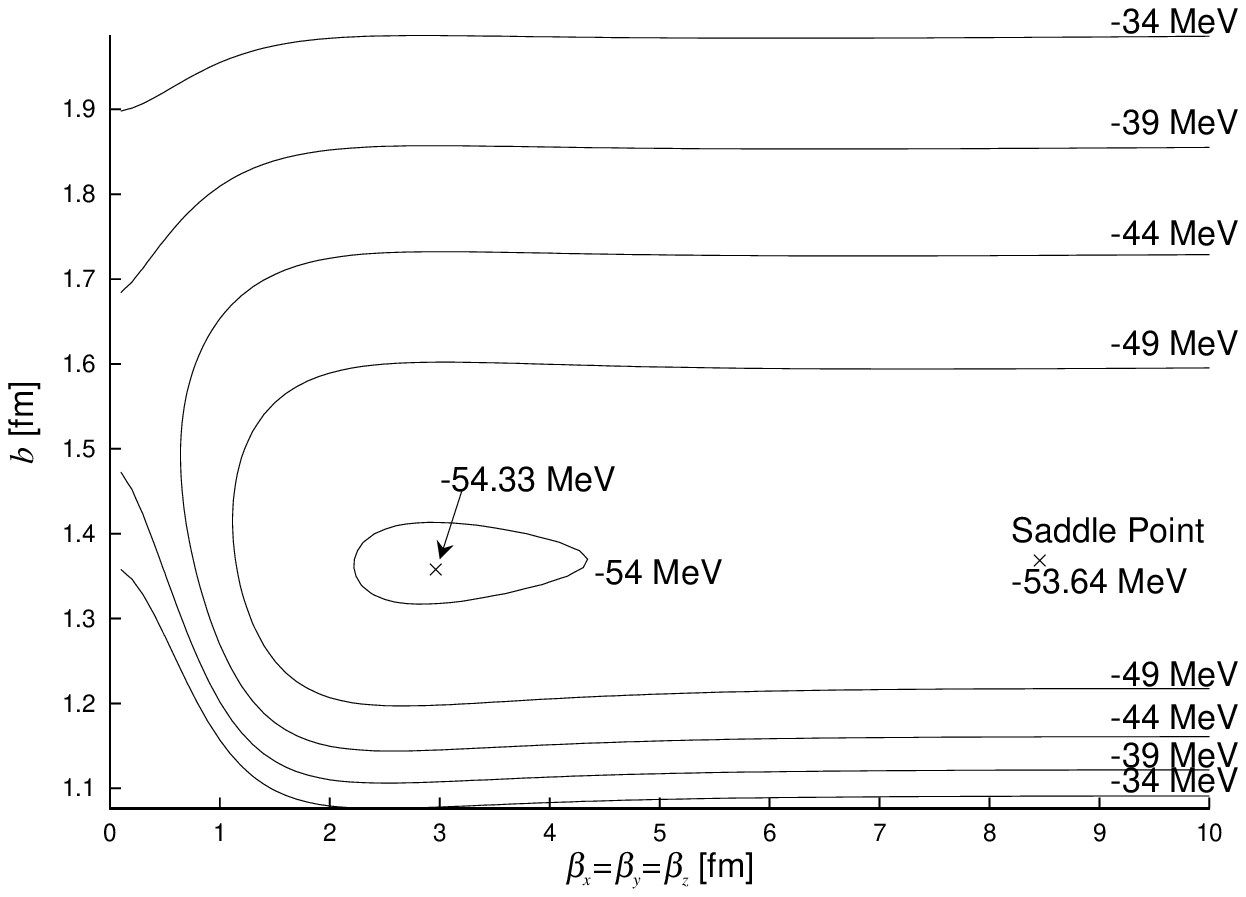}
Fig.1: Contour map of the energy surface of the spherical state for $M=0.56$ in the two parameter space, $b$ and $\beta_x=\beta_y=\beta_z$.
\end{figure}

\begin{figure}
\epsfxsize = \hsize
\epsfbox{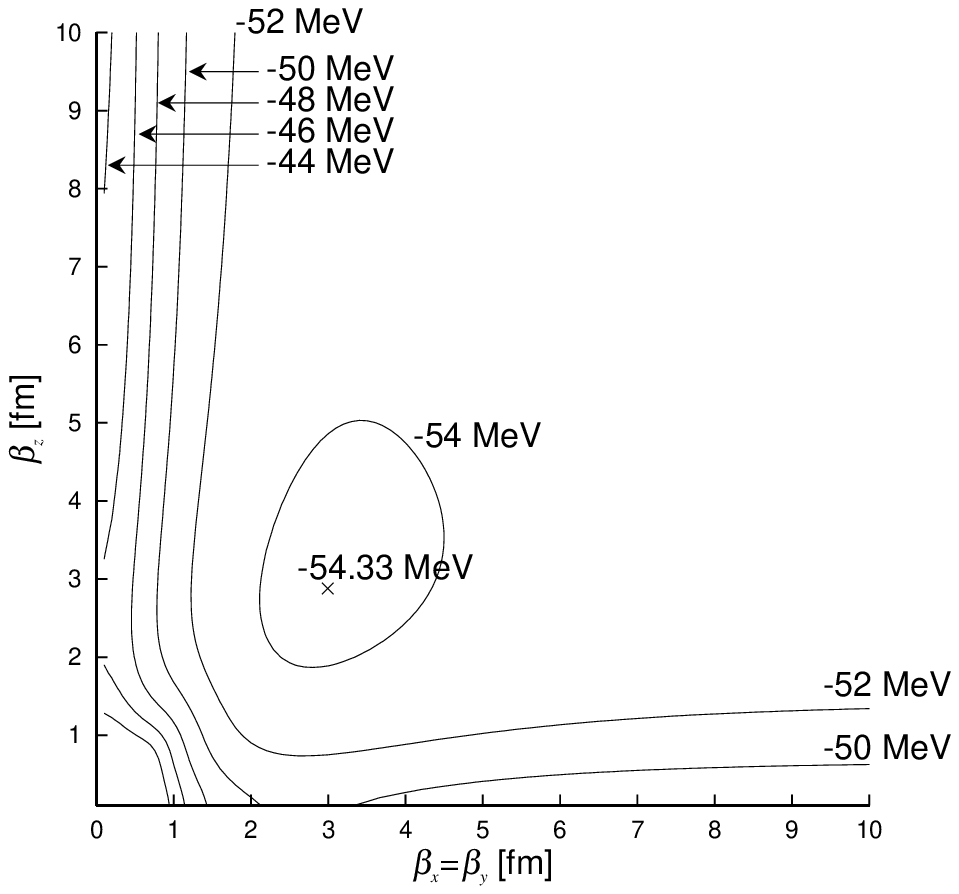}
Fig.2: Contour map of the energy surface of the deformed intrinsic state for $M=0.56$ in the two parameter space, $\beta_x(=\beta_y)$ and $\beta_z$.
\label{deformed56}
\end{figure}

\begin{figure}
\epsfxsize = \hsize
\epsfbox{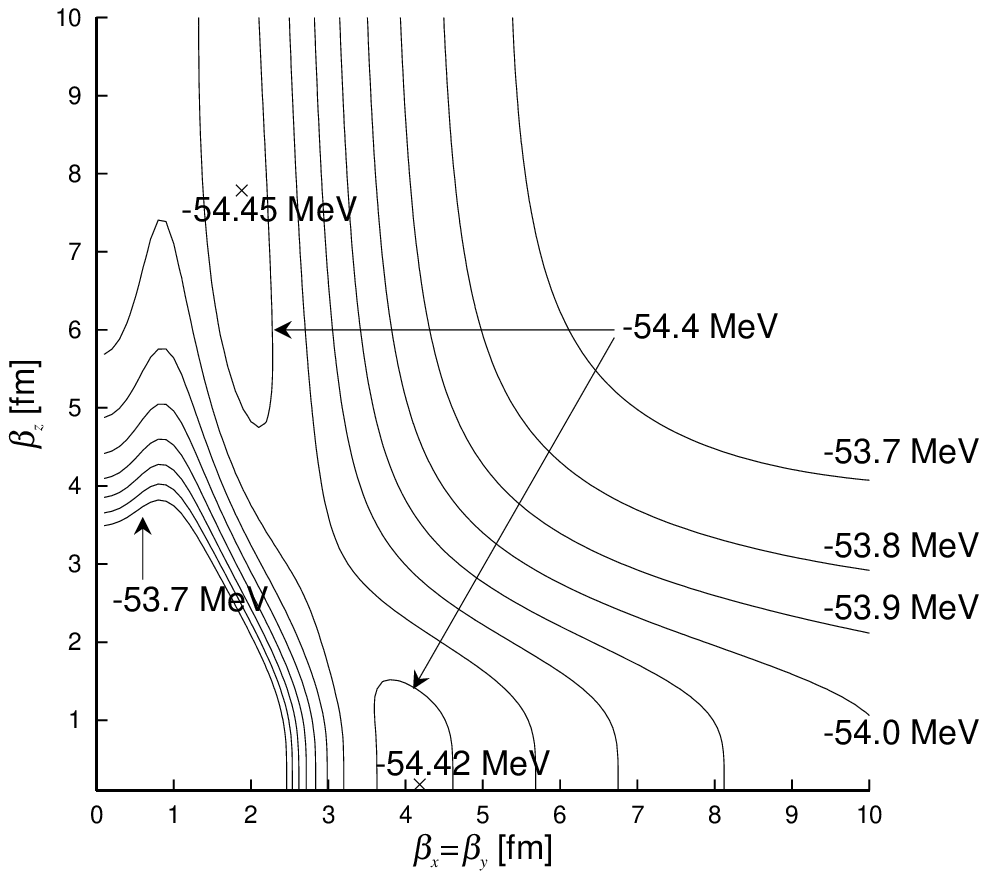}
Fig.3: Contour map of the energy surface of the $0^+$ state for $M=0.56$ in the two parameter space, $\beta_x(=\beta_y)$ and $\beta_z$.
\label{e0m56}
\end{figure}

\begin{figure}
\epsfxsize = \hsize
\epsfbox{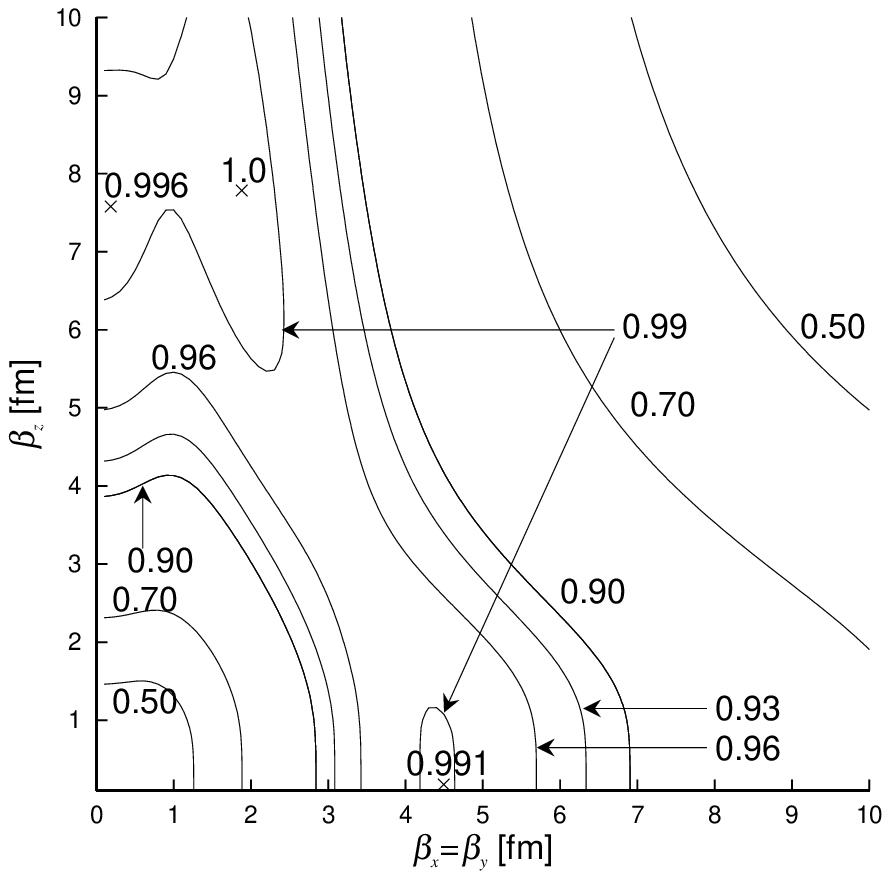}
Fig.4(a): Contour map of the squared overlap $O(\beta_x,\beta_z)$ between the $0^+$ wave function with $\beta_x=\beta_y=1.8$ fm, $\beta_z=7.8$ fm and the $0^+$ wave function with variable $\beta_x(=\beta_y)$ and  $\beta_z$. Numbers attached to the contour lines are squared overlap values.
\label{jjgcm0}
\end{figure}

\begin{figure}
\epsfxsize = \hsize
\epsfbox{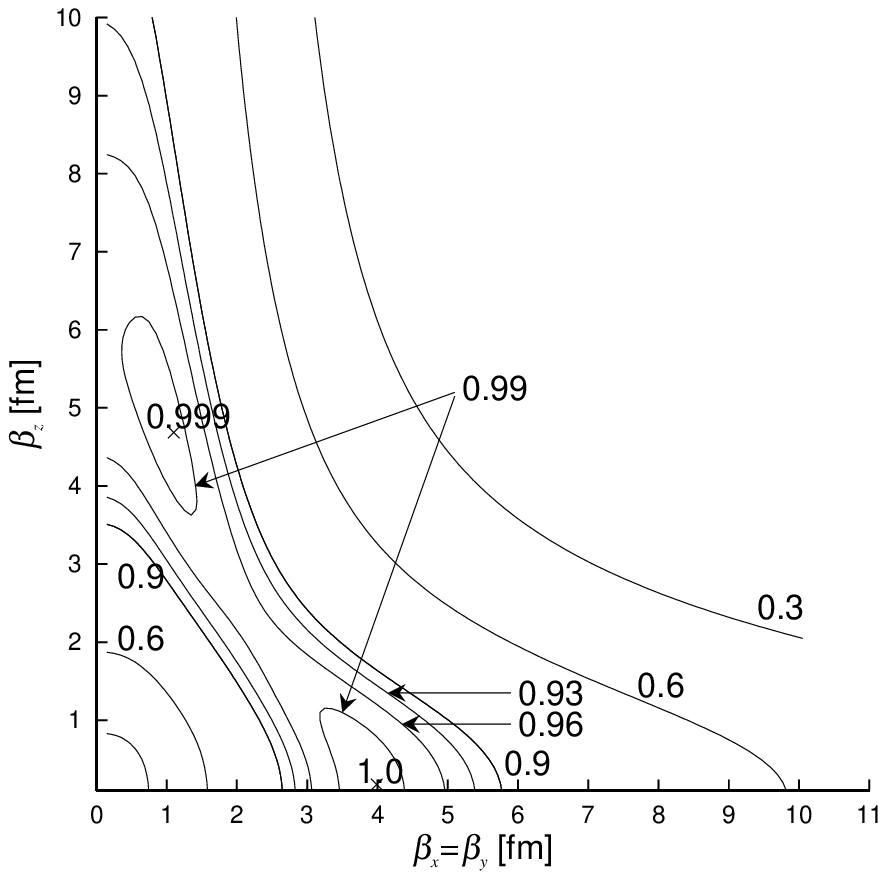}
Fig.4(b): Contour map of the squared overlap $O(\beta_x,\beta_z)$ between the $2^+$ wave function with $\beta_x=\beta_y=3.9$ fm, $\beta_z=0.0$ fm and the $2^+$ wave function with variable $\beta_x(=\beta_y)$ and $\beta_z$. Numbers attached to the contour lines are squared overlap values.
\label{jjgcm2}
\end{figure}

\begin{figure}
\epsfxsize = \hsize
\epsfbox{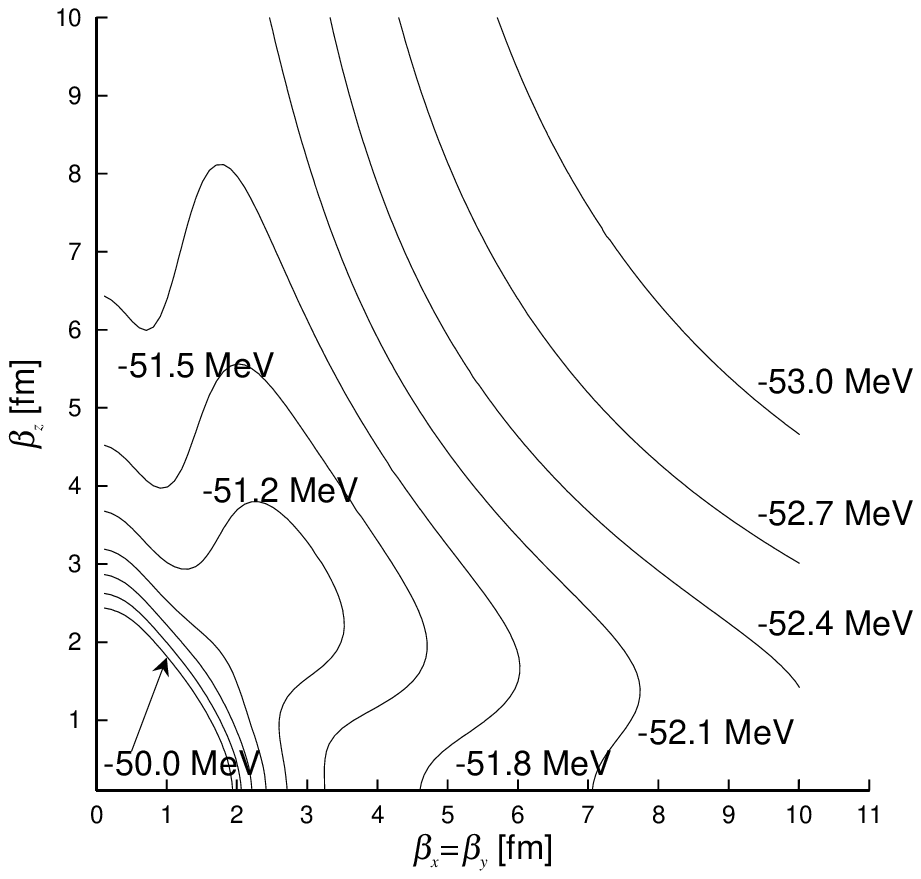}
Fig.5: Contour map of the  energy surface of the $2^+$ state for $M=0.56$ in the two parameter space, $\beta_x(=\beta_y)$ and $\beta_z$.
\label{e2m56}
\end{figure}

\begin{figure}
\epsfxsize = \hsize
\epsfbox{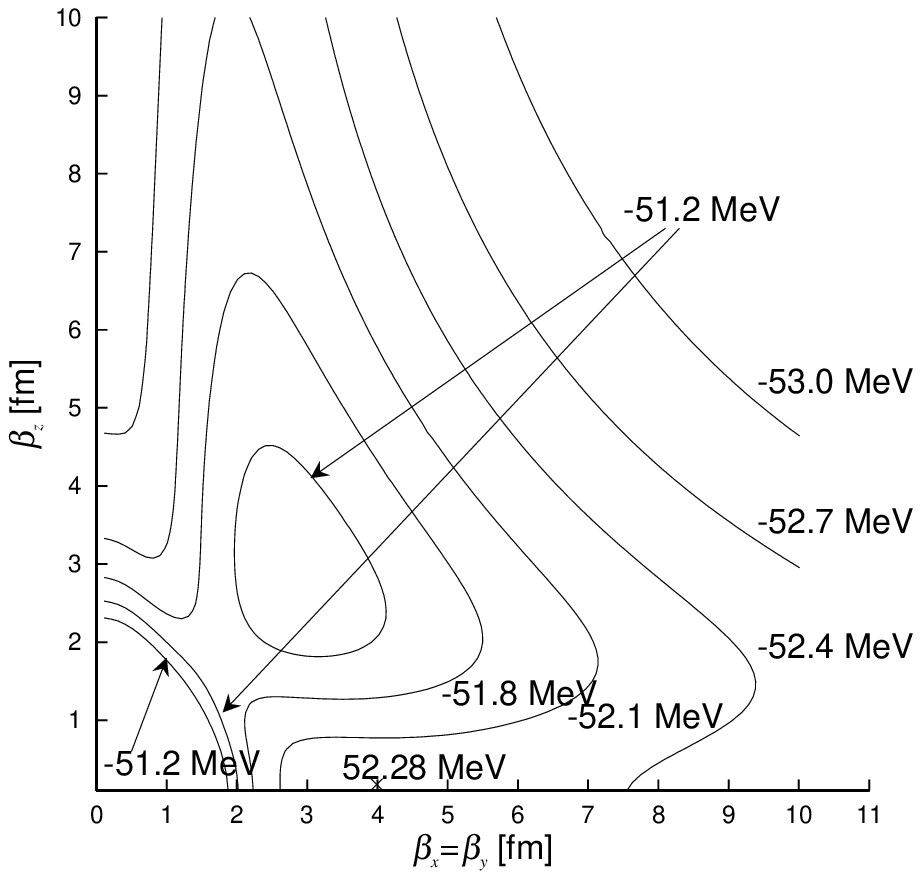}
Fig.6: Contour map of the energy surface of the $2^+$ state for $M=0.54$ in the two parameter space, $\beta_x(=\beta_y)$ and $\beta_z$.
\label{e2m54}
\end{figure}

\begin{figure}
\epsfxsize = \hsize
\epsfbox{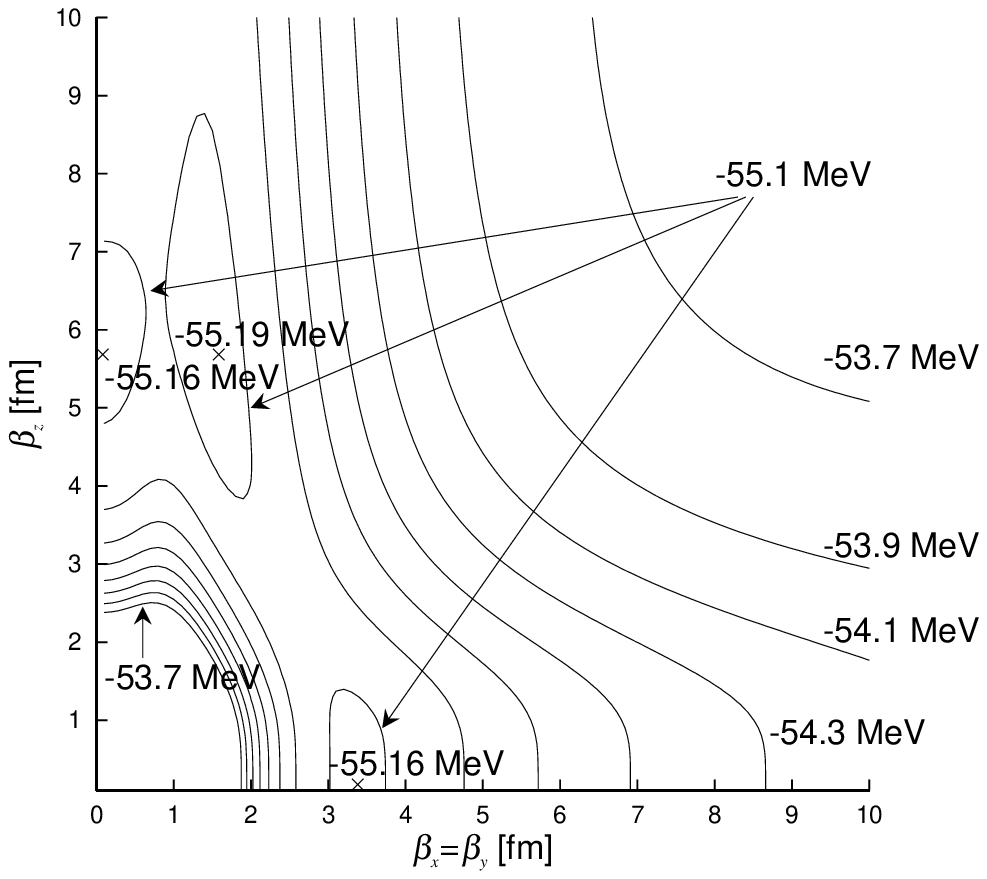}
Fig.7: Contour map of the energy surface of the $0^+$ state for $M=0.54$ in the two parameter space, $\beta_x(=\beta_y)$ and $\beta_z$.
\label{e0m54}
\end{figure}

\begin{figure}
\epsfxsize = \hsize
\epsfbox{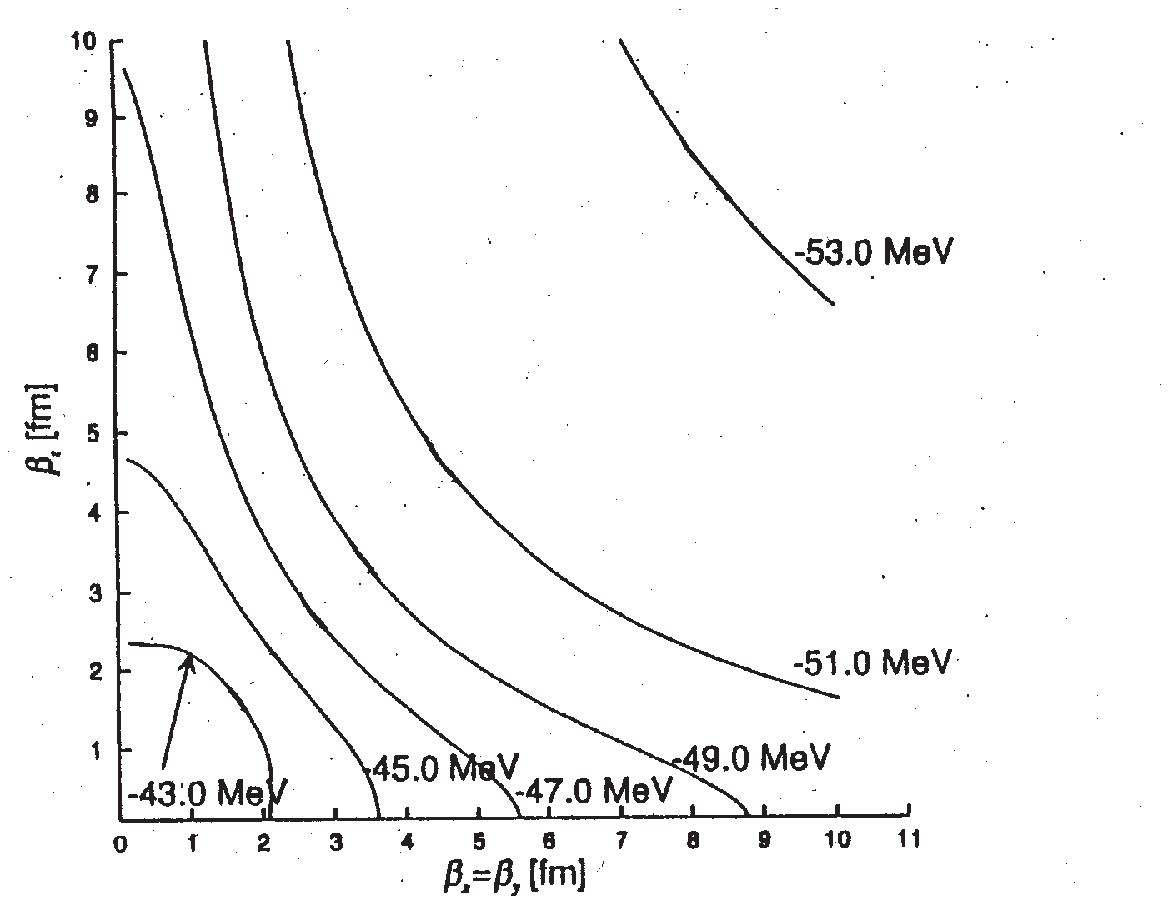}
Fig.8: Contour map of the energy surface of the $4^+$ state for $M=0.56$ in the two parameter space, $\beta_x(=\beta_y)$ and $\beta_z$.
\label{e4m56era}
\end{figure}

\begin{figure}
\epsfxsize = \hsize
\epsfbox{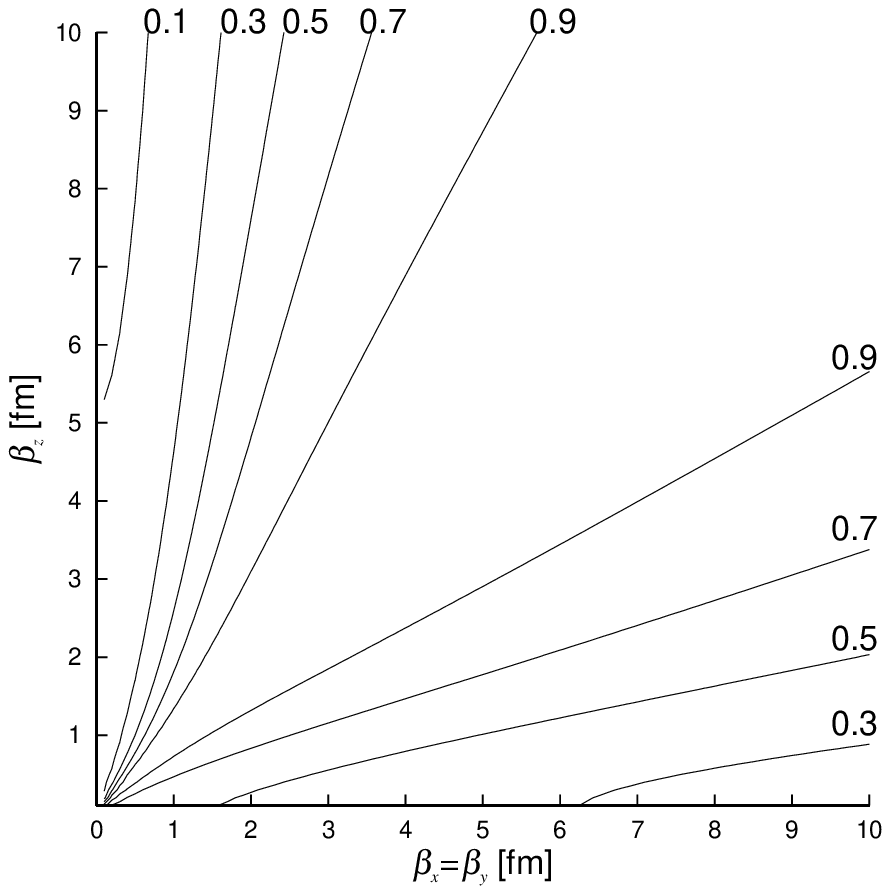}
Fig.9(a): Contour map of the squared amplitude of $J^+$ component contained in the deformed wave function in the two parameter space, $\beta_x(=\beta_y)$ and $\beta_z$ for $J=0$.
\label{amp0}
\end{figure}

\begin{figure}
\epsfxsize = \hsize
\epsfbox{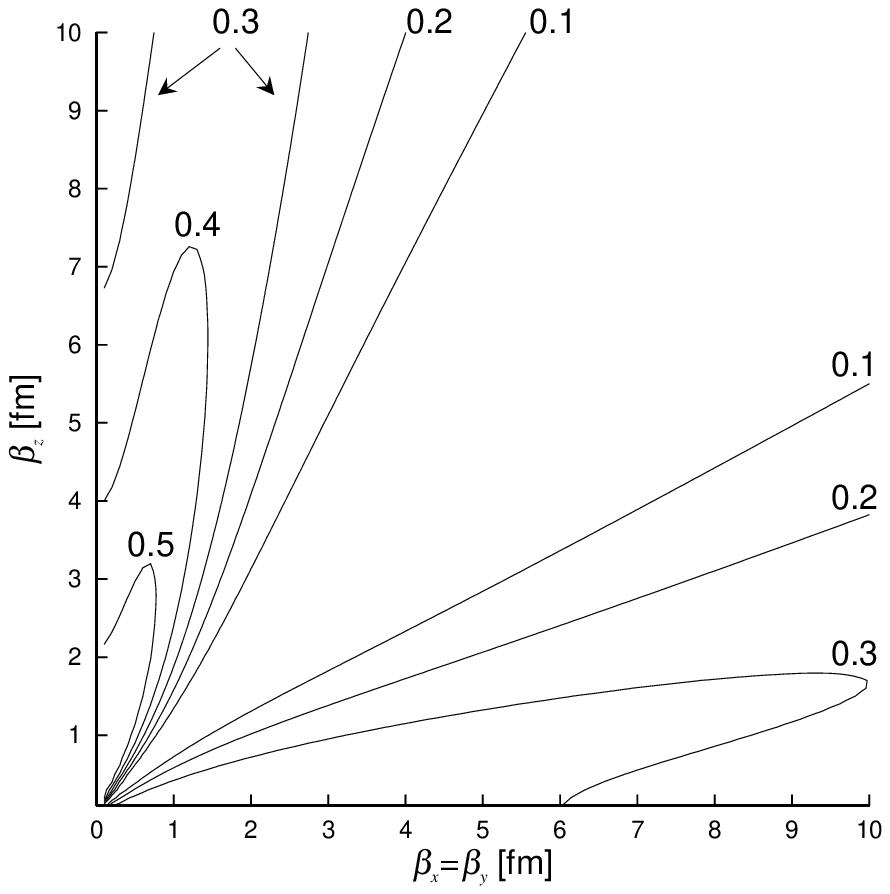}
Fig.9(b): Contour map of the squared amplitude of $J^+$ component contained in the deformed wave function in the two parameter space, $\beta_x(=\beta_y)$ and $\beta_z$ for $J=2$.
\label{a2}
\end{figure}

\begin{figure}
\epsfxsize = \hsize
\epsfbox{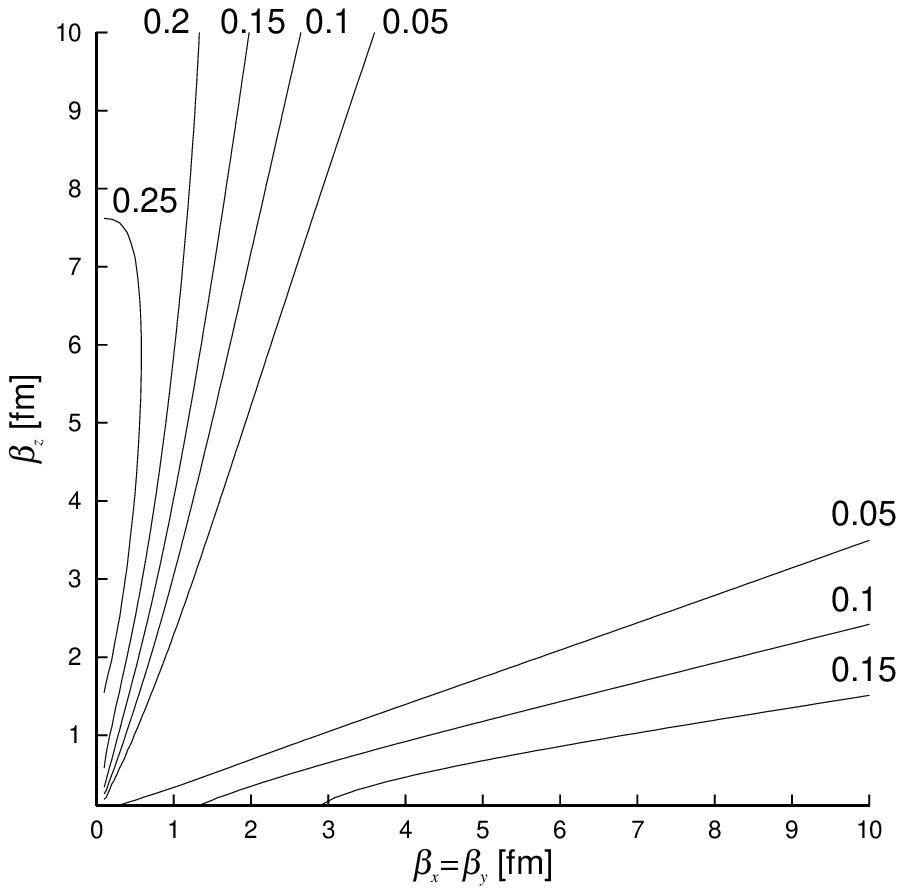}
Fig.9(c): Contour map of the squared amplitude of $J^+$ component contained in the deformed wave function in the two parameter space, $\beta_x(=\beta_y)$ and $\beta_z$ for $J=4$.
\label{a4}
\end{figure}

\begin{figure}
\epsfxsize = \hsize
\epsfbox{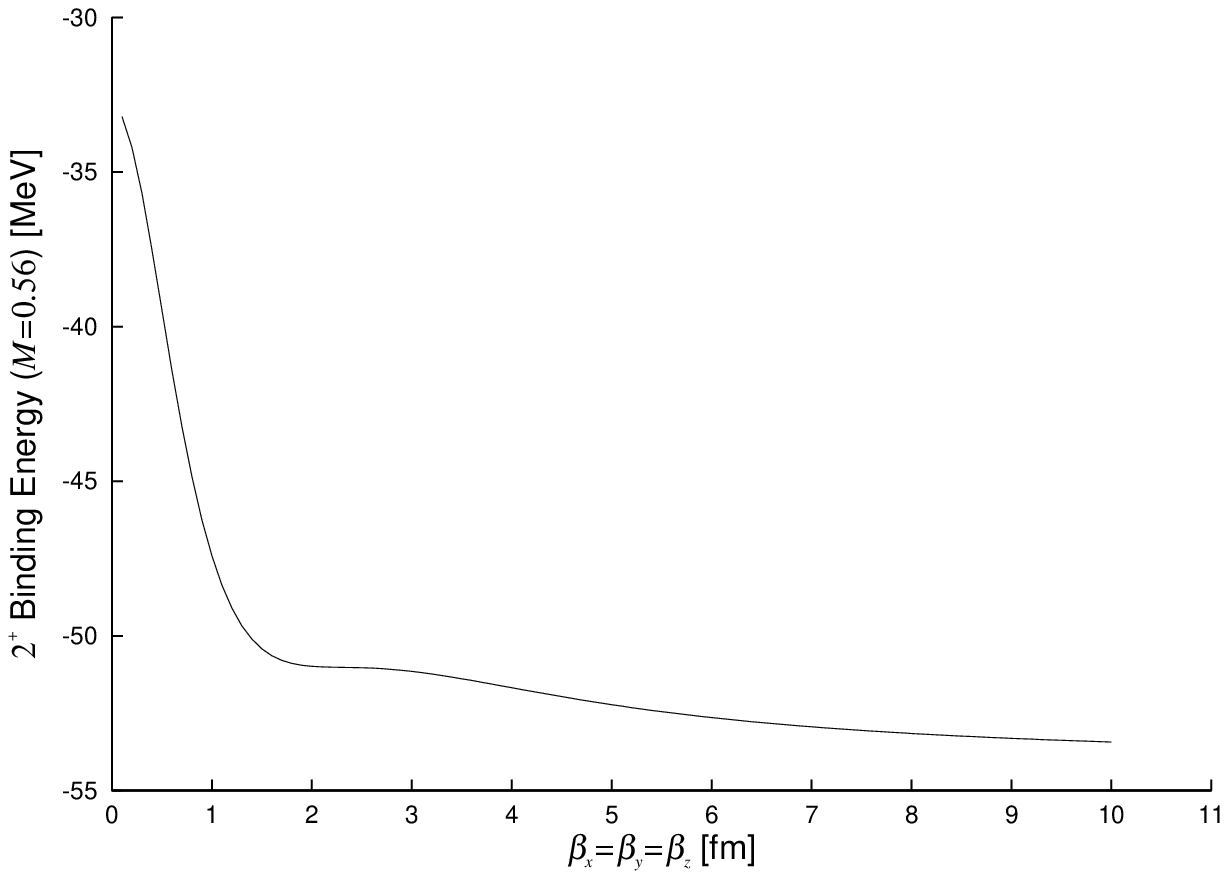}
Fig.10: Binding energy of the $2^+$ state on the $\beta_x=\beta_y=\beta_z$ line.
\label{ana2lim56}
\end{figure}

\begin{figure}
\epsfxsize = \hsize
\epsfbox{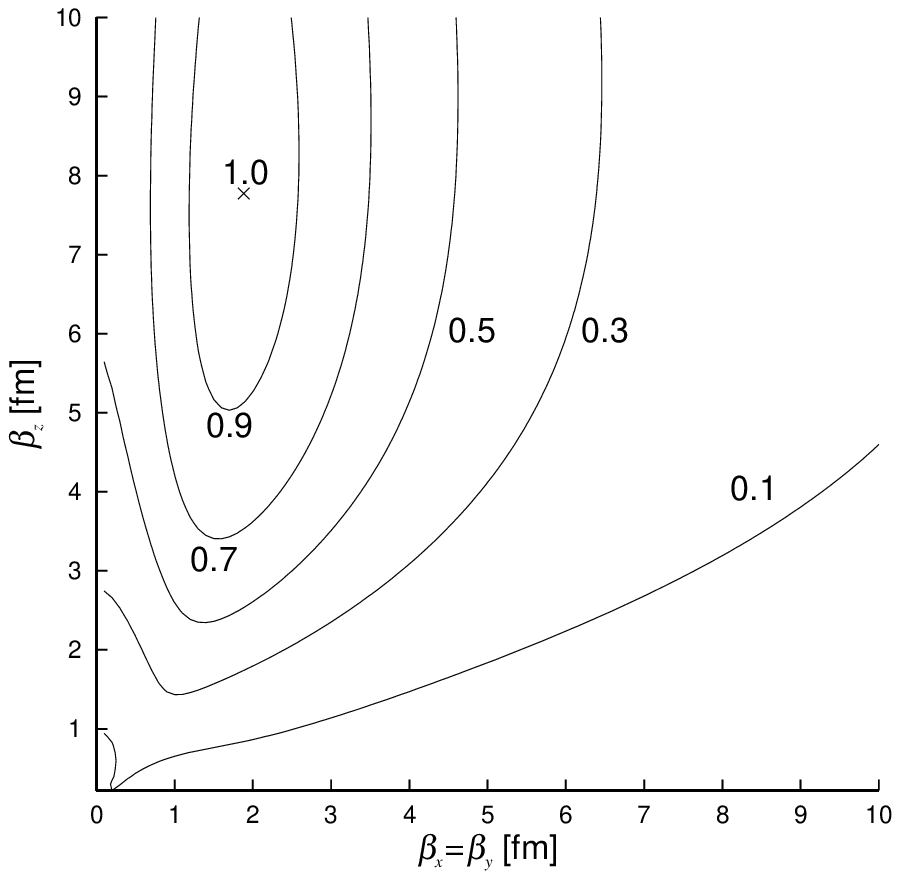}
Fig.11: Contour map of the squared overlap between the prolate intrinsic wave function with $\beta_x=\beta_y=1.8$ fm, $\beta_z=7.8$ fm and the intrinsic wave function with variable $\beta_x(=\beta_y)$ and $\beta_z$.
\label{gcmobpro}
\end{figure}